\documentclass[%
 reprint,
superscriptaddress,
 amsmath,amssymb,
 aps,
prb,
floatfix,
longbibliography
]{revtex4-2}

\usepackage[english]{babel}

\usepackage{ifpdf}
\usepackage{braket}
\usepackage{amsmath} 
\usepackage{amssymb}
\usepackage{amsfonts}
\usepackage{amsthm}
\usepackage{braket}
\usepackage{xcolor}
\usepackage[colorlinks=true,linkcolor=blue,citecolor=blue,breaklinks]{hyperref}
\usepackage{graphicx}
\usepackage{dcolumn}
\usepackage{bm}
\usepackage[normalem]{ulem}
\usepackage{verbatim}
\usepackage{algorithm}
\usepackage{algpseudocode}
\usepackage{graphicx}
\usepackage{afterpage}
\usepackage{graphicx}
\usepackage{placeins}

\usepackage[T1]{fontenc}
\usepackage{amsmath,amssymb,amsthm,mathtools}
\usepackage{microtype}
\usepackage{balance}

\definecolor{Zcolour}{rgb}{0.992, 0.588, 0.22}
\definecolor{purple}{rgb}{0.5,0,0.5}
\definecolor{brown}{rgb}{0.6,0.2,0}
\definecolor{dkgreen}{rgb}{0,0.5,0}

\newcommand{\rqA}{\overline{A}}
\newcommand{\rqDelta}{\overline{\Delta}}
\newcommand{\rqE}{\mathbb{E}}
\newcommand{\rqI}{\mathbb{I}}

\newcommand{\rqFrechet}[3]{D #1(#2)[#3]}
\DeclareMathOperator{\rqdiag}{diag}
\DeclarePairedDelimiter{\rqnorm}{\lVert}{\rVert}
\DeclarePairedDelimiter{\rqabs}{\lvert}{\rvert}

\newcommand{\rqscale}{\frac{\lambda_\epsilon}{\lambda}}

\newcommand{\sv}{^{\scriptscriptstyle(\mathrm{SV})}}

\newif\ifrqshownotes
\rqshownotestrue
\newcommand{\rqmark}[1]{\ifrqshownotes
  \textsuperscript{\hyperref[#1]{T\ref*{#1}}}\fi}
\newcounter{rqtranscriptionnote}

\allowdisplaybreaks[1]

\usepackage{mathtools}

\newtheorem{theorem}{Theorem}
\newtheorem{lemma}{Lemma}
\newtheorem{corollary}{Corollary}[theorem]

\begin{document}

\title{Double-Randomized Quantum Singular Value Transformations For Early Fault-Tolerant Quantum Computers}
\author{Joel Bierman}
\affiliation{Department of Electrical and Computer Engineering, North Carolina State University, Raleigh, NC 27606, USA}
\author{Pawan Poudel}
\affiliation{Department of Electrical and Computer Engineering, North Carolina State University, Raleigh, NC 27606, USA}
\author{Yuan Liu}%
 \email{q\_yuanliu@ncsu.edu}
\affiliation{Department of Electrical and Computer Engineering, North Carolina State University, Raleigh, NC 27606, USA}
\affiliation{Department of Computer Science, North Carolina State University, Raleigh, NC 27606, USA}
\affiliation{Department of Physics and Astronomy, North Carolina State University, Raleigh, NC 27607, USA}
\date{\today}

\date{\today}
\begin{abstract}
    The Quantum Singular Value Transformation (QSVT) is an algorithmic framework which encompasses a wide range of problems, including Hamiltonian dynamics, ground state preparation, Gibbs state preparation, amplitude amplification, and matrix inversion. For many areas of scientific interest such as quantum chemistry and condensed matter, the circuit depth of the block-encoding subroutine in QSVT can be formidable, limiting the scope of problems to which QSVT can be applied on early fault-tolerant hardware. We split linear operators $A$ into a core operator $A_0$ and a tail operator $A_\epsilon$, block-encoding the normalized sum of $A_0$ and the average of $n_T$ randomly sampled terms from $A_\epsilon$. We show that using randomized block-encodings of this form allows one to lower the circuit depth of QSVT while incurring a diamond norm error that scales as $\mathcal{O}(C_P\delta^2/n_T$), where $C_P$ is a prefactor which depends on the QSVT polynomial $P$ and $\delta$ is the ratio of the $\ell_1$ norms of the coefficient vectors of $A_\epsilon$ and $A$. We then go on to show that if one samples QSVT circuits from a joint probability distribution of both random block-encodings and random polynomials (Double-Randomized QSVT), then the circuit depth savings of these two randomization techniques combine multiplicatively, with only additive error contributions. We numerically calculate error bounds for long-range Heisenberg and iron-sulfur complex Hamiltonians, showing that for a diamond norm error bound of $10^{-3}$, RBE-QSVT can reduce the number of Pauli terms needed for the block-encoding by an order of magnitude or more in several cases. This paves the way for demonstrations of QSVT on early fault-tolerant hardware.
\end{abstract}

\maketitle

\section{Introduction}

The improvement of quantum computing hardware has accelerated in recent years, with below-threshold error correction~\cite{acharya_quantum_2025}, magic state distillation~\cite{sales_rodriguez_experimental_2025}, and error-corrected/detected quantum algorithms being demonstrated on hardware within the past few years~\cite{yamamoto_demonstrating_2024, perlin_fault-tolerant_2026, dasu_computing_2026}. Despite these impressive advances, we are still seemingly quite far away from having computers that can run the computations involving upwards of billions of quantum operations on hundreds of logical qubits that would be needed to solve problems that classical computers cannot~\cite{reiher_elucidating_2017, kivlichan_improved_2020}. These resource estimations typically assume the use of fault-tolerant algorithms with deep circuit depths. So-called "NISQ" algorithms such as the variational quantum eigensolver (VQE)~\cite{peruzzo_variational_2014, tilly_variational_2022, Grimsley2019} and sample-based diagonalization (SQD)~\cite{robledo-moreno_chemistry_2025, yu_quantum-centric_2025, piccinelli_quantum_2025, QBE_SQD} have been investigated as a means of solving problems with shallow circuits. However, these heuristic methods often lack rigorous performance guarantees, making their near-term utility uncertain.

So-called "early fault-tolerant" algorithms~\cite{katabarwa_early_2023, dong_ground-state_2022, wang_accelerated_2019} are typically constructed in such a way that rigorous runtime and error bounds can be proven in a similar fashion as their fully fault-tolerant counterparts while also reducing circuit depth. The trade-off typically made here is shallower circuits at the expense of worse asymptotic runtime scaling due to a larger sampling requirement.

Randomized compiling~\cite{hastings_turning_2016, campbell_shorter_2017, maupin_solovay-kitaev_2026} and randomized algorithms~\cite{wan_randomized_2022, wang_randomized_2025, tosta_randomized_2024, kivlichan_phase_2019, nakaji_high-order_2024, campbell_random_2019, noauthor_qubit-efficient_nodate, liu_ensemble-based_2026, martyn_halving_2025} often naturally exhibit these trade-offs in similar spirit to Monte Carlo algorithms \cite{kalos2008monte,gubernatis2016quantum}, making them compelling candidates for the early fault-tolerant era. Randomized compiling has, for example, been well-established as a means of error-mitigation.~\cite{temme_error_2017, suzuki_quantum_2022, liu_ensemble-based_2026} Indeed, it is common for randomized versions of algorithms to take approximation errors from $\mathcal{O(\epsilon)}$ to $\mathcal{O}(\epsilon^2)$~\cite{campbell_shorter_2017, martyn_halving_2025}
.
Quantum Signal Processing (QSP) and Quantum Singular Value Transformations (QSVT)~\cite{low2016methodology,low_optimal_2017,noauthor_grand_nodate, gilyen_quantum_2019, joven_scalable_2026, motlagh_generalized_2024, noauthor_doubling_nodate, rossi_multivariable_2022, dong_infinite_2024, lu_quantum_2026} are highly versatile fault-tolerant algorithmic frameworks which encompass a wide range of important scientific problems such as Hamiltonian dynamics~\cite{martyn_efficient_2023, low_optimal_2017, noauthor_grand_nodate, gilyen_quantum_2019, noauthor_doubling_nodate, motlagh_generalized_2024} and ground state preparation of quantum many-body system.~\cite{lin_optimal_2020, Lin2020nearoptimalground,dong2024multi, noauthor_grand_nodate}. The main barrier to implementing QSVT on near-term hardware is the cost of the block-encoding itself. For example, writing $A$ as a linear combination of $N$ unitaries, one can block-encode $A$ using a method such as linear combination of unitaries (LCU).~\cite{childs_hamiltonian_nodate} For problems such as quantum chemistry or other quantum many-body physics problems, $N$ can scale polynomially with the number of qubits $n$, bringing the total circuit depth to $\mathcal{O}(poly(n)d)$.

To-date, works on randomized variants of QSVT have been scarce compared to~\emph{e.g.}, randomized Trotterization methods~\cite{campbell_random_2019, nakaji_high-order_2024, ouyang_compilation_2020, piccinelli_quantum_2025}. Ref.~\cite{wang_randomized_2025} established a single-ancilla randomized QSVT method which involves $\mathcal{O}(1/\epsilon^2)$ calls to circuits with depth $\tilde{\mathcal{O}}(\lambda^2d^2)$ for operators of $1$-norm $\lambda$. Ref.~\cite{martyn_halving_2025} showed that one can use an effective QSVT degree $d_{eff}\approx\frac{d}{2}+\mathcal{O}(1)$ without sacrificing accuracy simply by sampling from an ensemble of polynomial. Ref.~\cite{tosta_randomized_2024} takes a conceptually similar approach as Ref.~\cite{martyn_halving_2025}, but samples from individual Chebyshev polynomials according to their coefficients. Ref.~\cite{paganelli_when_2026} showed that by sparsifying a block-encoded Hamiltonian $H$ using SparSto~\cite{ouyang_compilation_2020}, one can lower the block-encoding cost of QSVT while incurring operator norm error $\mathcal{O}(d\sqrt{\mathrm{Var_{coeff}}[H]})$, where $\mathrm{Var_{coeff}}[H]$ is a measure of the variance of the coefficient vector of the randomly sampled operators. Ref.~\cite{liu_ensemble-based_2026} showed that averaging over ensembles of noisy QSP circuits can mitigate coherent over or under-rotation errors. Ref.~\cite{maciel_robustness_2026} studied the robustness of QSP with respect to errors in the block-encoded operator and formulated algorithms which tailor the choice of polynomial to a given error.

In this work, we present a randomized QSVT method based on randomized block-encodings of random perturbations of a linear operator $A_0$ approximating the target operator $A$. That is, we express $A$ as a linear combination of operators with unit norm, partition it into a "core" $A_0$ and a "tail" $A_\epsilon$ comprised of $L_0$ and $L_\epsilon$, respectively and coefficient vector norms $\lambda_0$ and $\lambda_\epsilon$, respectively. Each QSVT circuit run performs a polynomial transformation of an operator comprised of the sum of $A_0$ and an operator proportional to the average of $n_T$ randomly sampled terms from $A_\epsilon$. We show that one can bound the diamond norm error of this randomized QSVT method as $\tfrac{C_P}{n_T}\left(\frac{\lambda_\epsilon}{\lambda}\right)^2$, where $\lambda$ is the norm of the coefficient vector of $A$ and $C_P$ is a prefactor which depends on the choice of polynomial. We then show by combining this randomized block-encoding method with the random polynomial sampling method of Ref.~\cite{martyn_halving_2025} in a joint distribution, one can inherit the circuit reduction benefits of both randomization techniques with error bounded by the sum of that of each method individually. We dub this method "double-randomized QSVT" (DR-QSVT). Importantly, the circuit depth of both methods scales as $\mathcal{O}((L_0+n_T) d)$, avoiding the $\mathcal{O}(d^2)$ scaling of single-ancilla randomized QSVT~\cite{wang_randomized_2025} while still achieving a lower circuit depth than conventional QSVT.

We note that the method of randomization used in this work is conceptually similar to that of Ref.~\cite{paganelli_when_2026}. We note the following important key differences of our work:
\begin{enumerate}
    \item Whereas Ref.~\cite{paganelli_when_2026} uses an ensemble of QSVT circuits such that each circuit layer uses an independently-sampled block-encoded operator, we take the block-encoding to be that of the same randomly-sampled operator at each layer within a given circuit run. This lets us establish $\left(\frac{\lambda_\epsilon}{\lambda}\right)^2$ error scaling using existing mixing lemmas as opposed to triangle inequality arguments. Furthermore, we can characterize the degree to which randomization is beneficial on a polynomial by polynomial basis through the prefactor $C_P$.
    \item We do not use SparSto-style randomization~\cite{ouyang_compilation_2020}, which allows us to make the number of Pauli terms in each block-encoding a constant as opposed to a random variable.
    \item We derive error bounds for the case where we perform joint sampling of both random polynomials and random block-encoded operators (double-randomized QSVT).
\end{enumerate}

The rest of this work is organized as follows. In Sec.~\ref{sec: background} we briefly summarize randomized compiling and QSVT. In Sec.~\ref{sec: rand qsvt method} we present the randomized QSVT method of this work, including error bounds for QSVT with randomized block-encodings (RBE-QSVT) and double-randomized QSVT (DR-QSVT). Numerical examples are given in Sec.~\ref{sec:numerical} for spin lattice models and quantum chemistry problems with representative applications. In Sec.~\ref{sec: conclusions} we conclude the work.
\section{Background}\label{sec: background}

Here we briefly summarize key prerequisite information for the rest of the work.

\subsection{QSVT}

On a high-level QSVT~\cite{noauthor_grand_nodate, gilyen_quantum_2019, joven_scalable_2026} is a method which takes as an input block-encodings of operators $A\in\mathbb{C}^{m\times n}$ such that $\left\|A\right\|\leq 1$:
\begin{equation}
    \mathrm{Be}\left[A\right] = \begin{bmatrix}
        A & *\\
        * & *
    \end{bmatrix}
\end{equation}and outputs block-encodings of polynomial transformations $P^{\sv}(A)$ of $A$:
\begin{equation}
    \mathrm{Be}\left[P^{\sv}(A)\right] = \begin{bmatrix}
        P^{\sv}(A) & *\\
        * & *
    \end{bmatrix}.
\end{equation}$P^{\sv}(A)$ is defined as~\cite{gilyen_quantum_2019, noauthor_grand_nodate}:
\begin{equation}
    P^{\sv}(A) = \begin{cases}
        \displaystyle\sum_{k=1}^{\min(m,n)} P(\sigma_k)\ket{\lambda_k^L}\bra{\lambda_k^R} ~\text{if $P$ is odd}\\
        \displaystyle\sum_{k=1}^n P(\sigma_k)\ket{\lambda_k^R}\bra{\lambda_k^R} ~\text{if $P$ is even}
    \end{cases}
\end{equation}where $\sigma_k$ are the singular values of $A$, $\ket{\lambda_k^L}$ and $\ket{\lambda_k^R}$ are the left and right singular vectors of $A$, and $P$ is our target polynomial. The class of preparable polynomials $P(x)$ must obey some mild constraints. Most importantly, they must be of definite parity (even or odd) and satisfy $|P(x)|\leq 1$ on the interval $[-1,1]$. Block-encoding a degree-$d$ polynomial using QSVT requires $d$ calls to $\mathrm{Be}[A]$ interleaved between projector-controlled rotation gates. The circuit is of the form:
\begin{equation}\label{eq: QSVT even circuit}
        U^{\vec{\phi}} = \left[\prod_{k=1}^{d/2}\Pi_{\phi_{2k-1}}\mathrm{Be}\left[A\right]^\dagger \tilde{\Pi}_{\phi_{2k}}\mathrm{Be}\left[A\right]\right]
\end{equation}for even $d$ and:
\begin{equation}\label{eq: QSVT odd circuit}
\begin{split}
        &U^{\vec{\phi}}= \tilde{\Pi}_{\phi_1}\mathrm{Be}\left[A\right]\\
        &\times\left[\prod_{k=1}^{(d-1)/2}\Pi_{\phi_{2k}}\mathrm{Be}\left[A\right]^\dagger \tilde{\Pi}_{\phi_{2k+1}}\mathrm{Be}\left[A\right]\right]
\end{split}
\end{equation}for odd $d$~\cite{noauthor_grand_nodate}. Here $\tilde{\Pi}_{\phi_j}$ and $\Pi_{\phi_j}$ are shorthand for the operations $e^{i\phi_j(2\tilde{\Pi}-I)}$ and $e^{i\phi_j(2\Pi-I)}$, where $\tilde{\Pi}$ and $\Pi$ are projectors onto the left and right singular vector spaces of $A$, respectively.  The phase factors $\left\{\phi_k\right\}$ can be efficiently computed classically as a pre-processing step~\cite{ni_fast_2024, dong_efficient_2021, skelton_hitchhikers_2025}. It is often the case that what we want is the real part of the polynomial and we consider the imaginary part to be extraneous. In this case we can perform a linear combination of two QSVT circuits with one additional ancilla qubit as:~\cite{gilyen_quantum_2019}
\begin{equation}\label{eq: re QSVT circuit}
    U_{\mathrm{Re}}^{\vec{\phi}} = (H\otimes\mathbb{I})(\ket{0}\bra{0}\otimes U^{\vec{\phi}}+\ket{1}\bra{1}\otimes U^{-\vec{\phi}})(H\otimes\mathbb{I})
\end{equation}where $H$ is the Hadamard gate in this context. We would then obtain the block-encoded polynomial by post-selecting on the $\ket{0}$ state of the ancilla as well as the ancillas already present in the QSVT circuits.

\subsection{Randomized Compiling}

Naively, one might be tempted to think that the concept of randomizing quantum circuit does not make much sense. After all, we already spend a lot of resources suppressing other sources of randomization such as gate errors. Why purposefully inject randomization? Would we not we make the circuit less accurate?

This intuition turns out to be incorrect in general. This (hypothetical) misconception stems from the unrealistic assumption that we had access to the exact unitary $U$ that we wish to run in the first place. In reality, we almost always have to resort to $\epsilon$-approximations $U_\epsilon$ of $U$. That is, if we have two quantum channels $\mathcal{U}(\rho) = U\rho U^\dagger$ and $\mathcal{U}_\epsilon(\rho) = U_\epsilon\rho U_\epsilon^\dagger$, then:
\begin{equation}
    \left\|\mathcal{U} - \mathcal{U}_\epsilon\right\|_\diamond \leq \epsilon.
\end{equation}The key insight of randomized compilation (and by extension many randomized algorithms) is that instead of implementing $U_\epsilon$, we introduce classical mixing, randomly sampling unitaries $U_j$ with probability $p_j$ from some pre-determined ensemble. If we choose this ensemble judiciously, then we can bound the diamond norm error of the resulting channel~\cite{campbell_shorter_2017, hastings_turning_2016}:
\begin{equation}
    \mathcal{M}(\rho) = \sum_jp_jU_j\rho U_j^\dagger
\end{equation}by
\begin{equation}
    \left\|\mathcal{M} - \mathcal{U}\right\|_\diamond = \mathcal{O}(\epsilon^2).
\end{equation}The diamond norm of a quantum channel $\Phi$ acting on a Hilbert space $\mathcal{H}$ is defined as:
\begin{equation}
    \left\|\Phi\right\|_\diamond = \sup_{\rho}\frac{\left\|(\Phi\otimes\mathcal{I}_n)(\rho)\right\|_1}{\left\|\rho\right\|_1}
\end{equation}where $n = \mathrm{dim}(\mathcal{H})$ and $\left\|\cdot\right\|_1$ is the trace norm defined as $\mathrm{tr(\sqrt{X^\dagger X})}$ for operator $X$. This quadratic suppression in error is the main finding of Hastings~\cite{hastings_turning_2016} and Campbell~\cite{campbell_shorter_2017}. Typically, $\left\{U_j\right\}$ are chosen to be easier to prepare than $U$.

\section{Methods}\label{sec: rand qsvt method}

\subsection{QSVT With Randomized Block-Encodings}\label{sec: main method}
In this section we consider a type of randomization where we have some linear operator $A_0$ which approximates a full operator $A\in\mathbb{C}^{m\times n}$. The full operator is a sum of many terms with norm 1:

\begin{equation}\label{eq: full Hamiltonian}
    A = \sum_{\ell=1}^L a_\ell\hat{P}_\ell
\end{equation}where $\|\hat{P}_\ell\|=1~\forall~\ell$. We have written $\hat{P}_\ell$ as many linear operators of interest are often expressed in the Pauli operator basis, however we note that nothing in the construction of these methods requires the use of this basis. $\hat{P}_\ell$ could be any linear operator of unit norm. We also note that throughout this work, a $P$ without a hat denotes a polynomial, not a Pauli operator. For operators expanded in a unitary basis, the cost of block-encoding would scale linearly with $L$ using the standard LCU formalism. Suppose therefore that (for a fixed QSP degree) we only have the ability to block-encode an approximate operator $A_0$ which is a sum of a subset of the terms in $A$:
\begin{equation}\label{eq: A_0}
    A_0 = \sum_{\ell=1}^{L_0}a_\ell^{(0)}\hat{P}_\ell^{(0)}.
\end{equation}This leaves us with an error operator $A_\epsilon = A - A_0$:
\begin{equation}\label{eq: H_e}
    A_\epsilon = \sum_{\mu=1}^{L_\epsilon}a_\mu^{(\epsilon)}\hat{P}_\mu^{(\epsilon)}.
\end{equation}The idea is that we can consider all the terms in $A_0$ to be sampled with probability 1 and independently sample $n_T$ terms from $A_\epsilon$ (with replacement) according to their weights. That is, we sample $n_T$ terms from $A_\epsilon$ according to the distribution with probabilities $\omega_\mu$:
\begin{equation}
    \begin{split}
        &\omega_\mu = \frac{|a_\mu^{(\epsilon)}|}{\lambda_\epsilon}, \qquad S_\mu = \mathrm{sign}(a_\mu^{(\epsilon)})\hat{P}_\mu^{(\epsilon)},
    \end{split}
\end{equation}where $\mathrm{sign}(z) = z/|z|$, $\lambda = \sum_{j=1}^L|a_j|$, and $\lambda_\epsilon = \sum_{\ell=1}^{L_\epsilon}|a_\ell^{(\epsilon)}|$. Let $\ell$ be shorthand for a vector of sampled operator indices, \emph{i.e.} $\ell = (\ell_1,\dots, \ell_t,\dots,\ell_{n_T})$. Then we sample operators $A_\ell$ with probability $p_\ell$ as:
\begin{equation}
        p_\ell = \prod_{t=1}^{n_T}\omega_{\ell_t}, ~~~A_\ell = \frac{A_0}{\lambda} + \frac{\lambda_\epsilon}{\lambda n_T}\sum_{t=1}^{n_T}S_{\ell_t}.
\end{equation}
We denote this ensemble by $\mathcal{D}_\epsilon$:
\begin{equation}
    \mathcal{D}_\epsilon = \left\{p_\ell: A_\ell\right\}_{\ell\in \mathcal{S}_T}, \qquad \mathcal{S}_T = \left\{1,\dots,L_\epsilon\right\}^{n_T}.
\end{equation}One can easily verify that the expectation value of this ensemble is $\mathbb{E}[A_\ell] = A/\lambda$. Throughout this manuscript, we will use the shorthand notation $\bar{A} = \mathbb{E}\left[A_\ell\right]$. The same sample $A_\ell$ is used in every block-encoding layer within a given circuit, but samples are independent across different circuit runs. For any degree-$d$ polynomial $P$ preparable by a QSP circuit, we can then randomly sample unitaries $U_\ell^{\vec{\phi}}$:
\begin{equation}\label{eq: QSVT even circuit ensemble}
        U_\ell = \left[\prod_{k=1}^{d/2}\Pi_{\phi_{2k-1}}\mathrm{Be}\left[A_\ell\right]^\dagger \tilde{\Pi}_{\phi_{2k}}\mathrm{Be}\left[A_\ell\right]\right]
\end{equation}for even $d$ and:
\begin{equation}\label{eq: QSVT odd circuit ensemble}
\begin{split}
        &U_\ell= \tilde{\Pi}_{\phi_1}\mathrm{Be}\left[A_\ell\right]\\
        &\times\left[\prod_{k=1}^{(d-1)/2}\Pi_{\phi_{2k}}\mathrm{Be}\left[A_\ell\right]^\dagger \tilde{\Pi}_{\phi_{2k+1}}\mathrm{Be}\left[A_\ell\right]\right]
\end{split}
\end{equation}for odd $d$. If what we want is the real part of the polynomial, then we would run an analog $U_{\mathrm{Re},\ell}^{\vec{\phi}}$ of the form in Eq.~\eqref{eq: re QSVT circuit}. Define:
\begin{equation}
    \Pi_P = \begin{cases}
        \Pi~\text{if P is even}\\
        \tilde{\Pi}~\text{if P is odd}
    \end{cases}.
\end{equation}The action of the completely positive  trace non-increasing map associated with this random sampling can be expressed as:
\begin{equation}
    \mathcal{R}(\ket{0}\bra{0}\otimes\rho) =\sum_{\ell\in\mathcal{S}_T}p_\ell \Pi_PU_\ell^{\vec{\phi}}(\ket{0}\bra{0}\otimes\rho) U_\ell^{\vec{\phi}\dagger}\Pi_P
\end{equation}The channel $\mathcal{U}$ corresponding to running QSVT with the exact block-encoding $A/\lambda$ is given by:
\begin{equation}
\begin{split}
    \mathcal{U}(\ket{0}\bra{0}\otimes\rho) = \Pi_PU^{\vec{\phi}}(\ket{0}\bra{0}\otimes\rho) U^{\vec{\phi}\dagger}\Pi_P\\
    \end{split}
\end{equation}where $U^{\vec{\phi}}$ is the QSVT circuit that block-encodes $P^{\sv}(A/\lambda)$. Let $\varrho = \ket{0}\ket{0}\bra{0}\bra{0}\otimes\rho$, where the first two registers are ancillas understood to not necessarily be the same size. The extension of these maps to real polynomials is given by:
\begin{equation}
    \begin{split}
        &\mathcal{R}_{\mathrm{Re}}(\varrho)\\ &=\sum_{\ell\in\mathcal{S}_T}p_\ell (\ket{0}\bra{0}\otimes\Pi_P)U_{\mathrm{Re},\ell}^{\vec{\phi}}\varrho U_{\mathrm{Re},\ell}^{\vec{\phi}\dagger}(\ket{0}\bra{0}\otimes\Pi_P)\\
    \end{split}
\end{equation}and
\begin{equation}
    \begin{split}
        \mathcal{U}_{\mathrm{Re}}(\varrho)= (\ket{0}\bra{0}\otimes\Pi_P)U_{\mathrm{Re}}^{\vec{\phi}}(\varrho) U_{\mathrm{Re}}^{\vec{\phi}\dagger}(\ket{0}\bra{0}\otimes\Pi_P).\\
    \end{split}
\end{equation}We write the corresponding maps that act on the input state $\rho$ as:
\begin{equation}
\begin{aligned}
 \mathcal{R}_P(\rho)&=\sum_\ell p_\ell P^{\sv}(A_\ell)\rho P^{\sv}(A_\ell)^\dagger,\\
 \mathcal{U}_P(\rho)&=P^{\sv}(\rqA)\rho P^{\sv}(\rqA)^\dagger.
\end{aligned}
\label{rq:eq:maps}
\end{equation}These maps can be interpreted to be comprised of QSVT circuits for either the real or complex polynomial cases depending on the context. We are interested in bounding the diamond norm of the difference between these two maps
   $ \rqnorm{\mathcal{R}_P-\mathcal{U}_P}_\diamond $.

\subsection{Double-Randomized QSVT}\label{sec: dr-qsvt}

DR-QSVT can be seen as an extension of RBE-QSVT with the incorporation of the polynomial-sampling method of Ref.~\cite{martyn_halving_2025}. This random polynomial sampling takes a degree-$d$ polynomial expanded in terms of Chebyshev polynomials $P(x) = \sum_{n=0}^{d} c_nT_n(x)$, partitions it into a core and a tail, and randomly samples polynomials which consist of the sum of the core and a term proportional to a random Chebyshev term from the tail:
\begin{equation}
    \begin{split}
        P_{core,k}(x) &= \sum_{n=0}^k c_nT_n(x)\\
        P_{tail,k}(x) &= \sum_{n=k+1}^d c_nT_n(x)
    \end{split}
    \end{equation}
    for $
        \epsilon_{P} = \sum_{n=k+1}^d|c_n|$, and sample polynomials from the distribution:
    \begin{equation}\label{eq: MR polynomial distribution}
        \mathcal{D}_{poly} = \left\{\pi_j = \frac{|c_j|}{\epsilon_{P}}: P_j(x) = \frac{P_{core,k}(x)}{\alpha} + \frac{c_j \epsilon_{P}}{\alpha|c_j|}T_j(x)\right\}.
    \end{equation}where
    \begin{equation}
        \alpha = \max_j\left\{\left\|P_{core,k}(x)+\frac{c_j\epsilon_P}{|c_j|}T_j(x)\right\|_\infty~|~x\in[-1,1]\right\}.
    \end{equation}We then sample operator-polynomial 2-tuples from the joint distribution:
    \begin{equation}\label{eq: joint distribution}
        \mathcal{D}_J = \left\{p_{\ell,j} = p_\ell\pi_j: (A_\ell, P_j(x))\right\}
    \end{equation} to generate the linear map:
    \begin{equation}
        \mathcal{J}_P(\rho) = \sum_{j,\ell}p_\ell\pi_jP_j^{\sv}(A_\ell)\rho P_j^{\sv}(A_\ell)^\dagger.
    \end{equation}
    We are interested in bounding the diamond norm error:
    \begin{equation}
        \left\|\mathcal{J}_P - \mathcal{U}_{P,\alpha}\right\|_\diamond
    \end{equation}where $\mathcal{U}_{P,\alpha} = \tfrac{1}{\alpha^2}\mathcal{U}_P$ and block-encodings of the polynomials $P_j^{\sv}(A_\ell)$ are generated with QSVT circuits directly analogous to those in Eq.~\eqref{eq: QSVT even circuit ensemble} and Eq.~\eqref{eq: QSVT odd circuit ensemble}.
\section{Error bounds}\label{sec: main result}
Throughout this work, we will use the following shorthand notation:
\begin{equation}
\begin{aligned}
 \Delta_\ell&=P^{\sv}(A_\ell)-P^{\sv}(\rqA),
 &\rqDelta&=\rqE[\Delta_\ell],\\
 q_P&=\rqnorm*{\rqE[\Delta_\ell^\dagger\Delta_\ell]},
 &b_P&=\rqnorm{\rqDelta}\\
 E_\ell&= A_\ell-\rqA\\
 V_L&= \mathbb{E}\left[E_\ell E_\ell^\dagger\right],&V_R&= \mathbb{E}\left[E_\ell^\dagger E_\ell\right]\\
 \nu&= \max\left\{\left\|V_L\right\|, \left\|V_R\right\|\right\}
\end{aligned}
\label{rq:eq:errors}
\end{equation}We now state the main results of our work that put bounds on the error of RBE-QSVT method described in Sec.~\ref{sec: main method} and DR-QSVT described in Sec.~\ref{sec: dr-qsvt}. These proofs rely on helper lemmas provided in Appendix~\ref{sec: helper lemmas}.

\begin{theorem}[QSVT with randomized block-encodings]
\label{rq:thm:main}
Fix an integer $n_T\geq 1$ and sample operators $A_\ell$ from $\mathcal{D}_\epsilon$. Let $g(x)$ be a function with compact support satisfying
\begin{equation}
 g(x)=P'(x)\quad\text{for }x\in[-1,1],\qquad \widehat{g}, t\widehat{g}\in L^1(\mathbb{R})
\label{rq:eq:g-assumptions}
\end{equation}where $\widehat{g}(t)$ is the Fourier transformation of $g(x)$. Define:
\begin{equation}
    \left\|P\right\| = \sup_{x\in[-1,1]}|P(x)|.
\end{equation}Then
\begin{equation}\label{rq:eq:main}
     \rqnorm{\mathcal{R}_P-\mathcal{U}_{P}}_\diamond
     \leq \frac{C_P}{n_T}\left(\rqscale\right)^2
\end{equation}
where
\begin{equation}
\rqnorm{\widehat{g}}_1 =\int_{\mathbb{R}}\rqabs{\widehat{g}(t)}\,dt,~~\rqnorm{\widehat{g^\prime}}_1 =\int_{\mathbb{R}}\rqabs{\widehat{g^\prime}(t)}\,dt
\label{rq:eq:fourier-moments}
\end{equation}and
\begin{equation}\label{eq: polynomial prefactor}
    C_P = \frac{1}{2\pi}\rqnorm{\widehat{g}}^2_1+\frac{||P||}{\sqrt{2\pi}}\rqnorm{\widehat{g^\prime}}_1.
\end{equation}
\end{theorem}
\begin{proof}
By Lemma~\ref{rq:lem:mixing}:
\begin{equation}
 \rqnorm{\mathcal{R}_P-\mathcal{U}_P}_\diamond
 \leq q_P+2\rqnorm{P}b_P.
\label{rq:eq:main-proof-mixing}
\end{equation}Lemma~\ref{lem: extended lemma} gives
\begin{equation}
 q_P\leq\frac{\nu}{2\pi} \rqnorm{\widehat{g}}_1^2,
 \qquad b_P\leq\frac{\nu}{2\sqrt{2\pi}}\rqnorm{\widehat{g^\prime}}_1.
\label{rq:eq:main-proof-bounds}
\end{equation}Let $t,u$ index a $draw$ and let $\ell_t, \ell_u$ index the $outcome$ of that draw for outcome vector $\ell$. Further let
\begin{equation}
    \bar{P} = \sum_{\mu=1}^{L_\epsilon}\omega_\mu S_\mu, \qquad Q_{\ell_t} = S_{\ell_t} - \bar{P}.
\end{equation}Then in the setup of Section~\ref{sec: main method}, we have
\begin{equation}
    E_\ell = \frac{\lambda_\epsilon}{n_T\lambda}\sum_{t=1}^{n_T}Q_{\ell_t},~\qquad  \mathbb{E}_\ell[Q_{\ell_t}] = 0.
\end{equation}For all $\ell$, draws $t$ and $u$ are independent for $t\neq u$, thus we have $\mathbb{E}_\ell[Q_{\ell_t}Q_{\ell_u}^\dagger] = \delta_{tu}\mathbb{E}_\ell[Q_{\ell_t}Q_{\ell_u}^\dagger]$. This allows us to write $V_L$ and $V_R$ as:
\begin{equation}
    \begin{split}
        &V_L = \frac{1}{n_T}\left(\frac{\lambda_\epsilon}{\lambda}\right)^2\left(\mathbb{E}_\mu\left[\hat{P}^{(\epsilon)}_\mu\hat{P}_\mu^{(\epsilon)\dagger}\right] - \bar{P}\bar{P}^\dagger\right)\\
        &V_R = \frac{1}{n_T}\left(\frac{\lambda_\epsilon}{\lambda}\right)^2\left(\mathbb{E}_\mu\left[\hat{P}^{(\epsilon)\dagger}_\mu\hat{P}_\mu^{(\epsilon)}\right] - \bar{P}^\dagger\bar{P}\right)
    \end{split}
\end{equation}where we have used the fact that $\mathbb{E}_\ell[Q_{\ell_t}Q_{\ell_t}^\dagger] = \mathbb{E}_\mu[\hat{P}_\mu^{(\epsilon)}\hat{P}_\mu^{(\epsilon)\dagger}]-\bar{P}\bar{P}^\dagger$ and $\mathbb{E}_\ell[Q_{\ell_t}^\dagger Q_{\ell_t}] = \mathbb{E}_\mu[\hat{P}_\mu^{(\epsilon)\dagger}\hat{P}_\mu^{(\epsilon)}] - \bar{P}^\dagger \bar{P}$. $V_L$ and $V_R$ can both be seen to be positive semidefinite. For example, for any $\ket{\psi}\in \mathbb{C}^m$, we have:
\begin{equation}
\begin{split}
\bra{\psi}V_L\ket{\psi} &= \sum_\ell p_\ell\bra{\psi}E_\ell E_\ell^\dagger\ket{\psi}\geq 0.
\end{split}
\end{equation}An analogous statement is true for $V_R$ for any $\ket{\psi}\in \mathbb{C}^n$. Thus, all the eigenvalues of $V_L, V_R$ are non-negative. To prove upper bounds on $\left\|V_L\right\|$ and $\left\|V_R\right\|$, we can observe that $\bar{P}^\dagger \bar{P}$ and $\bar{P}\bar{P}^\dagger$ are also both positive semidefinite. The fact that $\rqnorm{\mathbb{E}_\mu[\hat{P}_\mu^{(\epsilon)\dagger}\hat{P}_\mu^{(\epsilon)}]}\leq 1$ and $\rqnorm{\mathbb{E}_\mu[\hat{P}_\mu^{(\epsilon)}\hat{P}_\mu^{(\epsilon)\dagger}]}\leq 1$ therefore implies
\begin{equation}
    \|V_L\|,\ \|V_R\| \leq \frac{1}{n_T}
    \left(\frac{\lambda_\epsilon}{\lambda}\right)^2.
\end{equation}Then
\begin{equation}
\begin{split}
 &\rqnorm{\mathcal{R}_P-\mathcal{U}_{P}}_\diamond\leq \frac{C_P}{n_T}\left(\rqscale\right)^2
\end{split}
\end{equation}
\end{proof}
\begin{theorem}[Double-Randomized QSVT]\label{thm: DR-QSVT}
    Consider the operator sampling method of Sec.~\ref{sec: main method}. Let $\mathcal{D}_{poly} = \left\{\pi_j: P_j(x)\right\}$ be a distribution on polynomials allowed by QSVT and with the same parity such that $\mathbb{E}\left[P_j(x)\right] = \tfrac{1}{\alpha}P(x)~\forall~x\in[-1,1]$. Let $\mathcal{J}_P$ be the linear map associated with independent joint sampling from $\mathcal{D}_{poly}$ and $\mathcal{D}_\epsilon$:
    \begin{equation}\label{eq: joint QSVT CPTNI channel}
        \mathcal{J}_P(\rho) = \sum_{j,\ell}p_\ell\pi_jP_j^{\sv}(A_\ell)\rho P_j^{\sv}(A_\ell)^\dagger.
    \end{equation} We can bound the diamond norm error of this linear map as:
    \begin{equation}
        \left\|\mathcal{J}_P -\mathcal{U}_{P,\alpha}\right\|_\diamond \leq \tfrac{1}{\alpha^2}\left\|\mathcal{R}_P - \mathcal{U}_{P}\right\|_\diamond + \nu_{P}
    \end{equation}where:
    \begin{equation}
        \nu_P = \max_{x\in[-1,1]}\left\{\sum_j \pi_j |P_j(x) - \tfrac{1}{\alpha}P(x)|^2\right\}
    \end{equation}and $\left\|\mathcal{R}_P - \mathcal{U}_P\right\|_\diamond$ is as is in Theorem~\ref{rq:thm:main}.
\end{theorem}
\begin{corollary}[DR-QSVT with Chebyshev polynomial tail sampling]\label{cor: DR-QSVT MR sampling}
    Consider the method of sampling from $\mathcal{D}_{poly}$ as described in Ref.~\cite{martyn_halving_2025}. That is, we sample polynomials from the distribution in Eq.~\eqref{eq: MR polynomial distribution} and sample operator-polynomial 2-tuples $(A_\ell, P_j(x))$ from Eq.~\eqref{eq: joint distribution}. Then we can bound the diamond norm error by:
    \begin{equation}
        \left\|\mathcal{J}_P -\mathcal{U}_{P,\alpha}\right\|_\diamond \leq \tfrac{1}{\alpha^2}\left\|\mathcal{R}_P - \mathcal{U}_P\right\|_\diamond + \left(\frac{\epsilon_P}{\alpha}\right)^2
    \end{equation}with average circuit depth:
    \begin{equation}
        \mathbb{E}\left[depth\right] = \mathbb{E}_\ell[D_{BE}(A_\ell)]\sum_{j=k+1}^d\frac{|c_j|}{\epsilon_P}j
    \end{equation}where $\mathbb{E}_\ell[D_{BE}(A_\ell)]$ is the average block-encoding cost $D_{A_\ell}$ associated with block-encoding $A_\ell$ sampled from the ensemble $\mathcal{D}_\epsilon$ considered in Theorem.~\ref{rq:thm:main} and $\epsilon_P = \sum_{j=k+1}^{d}|c_j|$. For simplicity we have absorbed the cost of the projector-controlled rotation gates into $D_{BE}(A_\ell)$.
\end{corollary}
\begin{proof}The steps largely mirror the mixing lemma proofs~\cite{hastings_turning_2016, campbell_shorter_2017}. Let $D_j(x)=P_j(x)-\tfrac{1}{\alpha}P(x)$. Then:
\begin{equation}
\begin{aligned}
&\mathbb E_j\!\left[P_j(x)\rho P_j(x)^\dagger\right]\\
&\quad=\tfrac
{1}{\alpha^2}P(x)\rho P(x)^\dagger+\mathbb E_j\!\left[D_j(x)\rho D_j(x)^\dagger\right]
\end{aligned}
\label{eq:hr-handwritten-polynomial-expansion}
\end{equation}where we have used the facts that $\mathbb{E}\left[D_j(x)\right] = 0$ and that $\rho$ and $P(x)$ are constants to get rid of the cross terms. Thus
\begin{equation}
\begin{aligned}
&\mathcal J_P(\rho)-\mathcal U_{P,\alpha}(\rho)\\
&\quad=\mathbb E_\ell\!\Bigl[
    \tfrac{1}{\alpha^2}
    P^{\sv}(A_\ell)\rho P^{\sv}(A_\ell)^\dagger\\
&\qquad\quad
    +\mathbb E_j\!\bigl[
        D_j^{\sv}(A_\ell)\rho D_j^{\sv}(A_\ell)^\dagger
    \bigr]\Bigr]\\
&\qquad-\tfrac{1}{\alpha^2}
    P^{\sv}(\rqA)\rho P^{\sv}(\rqA)^\dagger.
\end{aligned}
\label{eq:hr-handwritten-channel-difference}
\end{equation}Let
\begin{equation}
\begin{aligned}
\mathcal V_P(\rho)
&=\mathbb E_\ell\!\left[
  \mathbb E_j\!\left[D_j^{\sv}(A_\ell)\rho D_j^{\sv}(A_\ell)^\dagger\right]
  \right]
\end{aligned}
\label{eq:hr-handwritten-variance-map}
\end{equation}We can then bound the diamond norm error by:
\begin{equation}
\begin{aligned}
&\left\lVert\mathcal J_P-\mathcal U_{P,\alpha}\right\rVert_\diamond\le
\tfrac{1}{\alpha^2}\left\lVert\mathcal R_P-\mathcal U_P\right\rVert_\diamond
+\left\lVert\mathcal V_P\right\rVert_\diamond
\end{aligned}
\label{eq:hr-handwritten-triangle-bound}
\end{equation}
By [\cite{watrous2018theory}, Theorem 3.39, Definition 3.37, and Definition 3.43], we have:
\begin{equation}
\begin{aligned}
    \left\lVert\mathcal V_P\right\rVert_\diamond
&=\left\|\mathcal{V}_P^\dagger(\mathbb{I})\right\|\\
&=\left\lVert
  \sum_\ell\sum_j p_\ell\pi_j
  D_j^{\sv}(A_\ell)^\dagger D_j^{\sv}(A_\ell)
  \right\rVert.\\
\end{aligned}
\end{equation}For all $\ell$ we can write:
\begin{equation}
\begin{aligned}
    &\sum_j\pi_j
  D_j^{\sv}(A_\ell)^\dagger D_j^{\sv}(A_\ell)\\
  &=\sum_a\left[\sum_j\pi_j|D_j(\sigma_{a,\ell})|^2\right]\ket{\lambda^{(R)}_{a,\ell}}\bra{\lambda^{(R)}_{a,\ell}}
\end{aligned}
\end{equation}where $\sigma_{a,\ell}$ and $\ket{\lambda_{a,\ell}^{(R)}}$ are the singular values and right singular vectors of $A_\ell$, respectively. We therefore have
\begin{equation}
\begin{aligned}
&\left\lVert\mathcal V_P\right\rVert_\diamond\\
&=\left\lVert\mathbb E_\ell\ \left[
  \sum_a\left[\sum_j\pi_j|D_j(\sigma_{a,\ell})|^2\right]\ket{\lambda^{(R)}_{a,\ell}}\bra{\lambda^{(R)}_{a,\ell}}
  \right]\right\rVert\\
&\leq\max_{x\in[-1,1]}\left\{\sum_j \pi_j |P_j(x) - \tfrac{1}{\alpha}P(x)|^2\right\}.
\end{aligned}
\label{eq:hr-handwritten-cp-bound}
\end{equation}
Thus
\begin{equation}
\begin{aligned}
&\left\lVert\mathcal J_P-\mathcal U_{P,\alpha}\right\rVert_\diamond\le
\frac{1}{\alpha^2}\left\lVert\mathcal R_P-\mathcal U_P\right\rVert_\diamond+\nu_P
\end{aligned}
\label{eq:hr-handwritten-joint-bound}
\end{equation}This proves Theorem~\ref{thm: DR-QSVT}. We now prove Corollary~\ref{cor: DR-QSVT MR sampling}. We specialize to sampling $A_\ell$ according to the set-up of Theorem~\ref{rq:thm:main}
and sampling $P_j(x)$ from Eq.~\eqref{eq: MR polynomial distribution}. Thus:
\begin{equation}
\begin{aligned}
&\sum_{j=k+1}^{d}\pi_j\left\lvert D_j(x)\right\rvert^2\\
&\quad=\frac{1}{\alpha^2}\left[\epsilon_P\sum_{j=k+1}^{d}|c_j|T_j(x)^2
  -|P_{\mathrm{tail},k}(x)|^2\right]\\
&\quad\le\left(\frac{\epsilon_P}{\alpha}\right)^2.
\end{aligned}
\label{eq:hr-handwritten-tail-variance}
\end{equation}
By Theorem~\ref{rq:thm:main} we have:
\begin{equation}
\begin{aligned}
&\rqnorm{\mathcal{R}_P-\mathcal{U}_{P}}_\diamond\leq \frac{C_P}{n_T}\left(\rqscale\right)^2
\end{aligned}
\label{eq:hr-handwritten-sobolev-bound}
\end{equation}where $C_P$ is as in Eq.~\eqref{eq: polynomial prefactor}. Thus:
\begin{equation}
\begin{aligned}
&\left\lVert\mathcal J_P-\mathcal U_{P,\alpha}\right\rVert_\diamond\leq \frac{C_P}{\alpha^2n_T}\left(\rqscale\right)^2+\left(\frac{\epsilon_P}{\alpha}\right)^2
\end{aligned}
\label{eq:hr-handwritten-specialized-joint-bound}
\end{equation}
The average circuit depth is furthermore given by:
\begin{equation}
\mathbb E_\ell\!\left[D_{\mathrm{BE}}(A_\ell)\right]
\sum_{j=k+1}^{d}\frac{|c_j|}{\epsilon_P}\,j.
\label{eq:hr-handwritten-average-depth}
\qedhere
\end{equation}
\end{proof}We note that maps of the form Eq~\eqref{rq:eq:maps} are not true quantum channels as they do not preserve the trace. That is, they are not completely positive trace-preserving (CPTP), but completely positive trace non-increasing (CPTNI). The bounds given in Thm.~\ref{rq:thm:main} and Thm.~\ref{thm: DR-QSVT} answer the question of "given an ensemble of randomized QSVT circuits, how close is the random block-encoded polynomial to the exact polynomial we want?". It does not necessarily answer the question "given an ensemble of randomized QSVT circuits, how close is the post-selected output quantum state to the exact one we want?". Thus, some care is required regarding how we interpret the previous theorems. We address the latter question in the following theorem.
\begin{theorem}[Error bounds on output postselected states]
Let $\mathcal{U}$ be a CPTNI QSVT map of the form in Eq.~\eqref{rq:eq:maps} and let $\mathcal{R}$ be a randomized CPTNI QSVT map of the form in Eq.~\eqref{rq:eq:maps} or Eq.~\eqref{eq: joint QSVT CPTNI channel}. Let $p =\mathrm{tr}\left(\mathcal{U}(\rho)\right)> 0$ be the post-selection probability of measuring the $\ket{0}$ block of the QSVT circuit. Further suppose $q = \mathrm{tr}
        \left(\mathcal{\mathcal{R}}(\rho)\right) > 0$. Let $\epsilon$ be an upper-bound on $\rqnorm{\mathcal{R}-\mathcal{U}}_\diamond$. Then:
\begin{equation}
    \left\|\frac{\mathcal{R}(\rho)}{q}-\frac{\mathcal{U}(\rho)}{p}\right\|_1 \leq \frac{2\epsilon}{p}.
\end{equation}
\end{theorem}
\begin{proof}Adding and subtracting $\mathcal{R}(\rho)/p$, we have:
\begin{equation}
    \begin{split}
        &\left\|\frac{\mathcal{R}(\rho)}{q}-\frac{\mathcal{U}(\rho)}{p}\right\|_1\\
        &\leq \left\|\frac{\mathcal{R}(\rho)}{q}-\frac{\mathcal{R}(\rho)}{p} + \frac{\mathcal{R}(\rho)}{p} - \frac{\mathcal{U}(\rho)}{p}\right\|_1\\
        &\leq \left\| \frac{\mathcal{R}(\rho)}{q} - \frac{\mathcal{R}(\rho)}{p}\right\|_1 + \left\|\frac{\mathcal{R}(\rho)}{p} - \frac{\mathcal{U}(\rho)}{p}\right\|_1.
    \end{split}
\end{equation}Since $\mathcal{R}(\rho)$ is positive semidefinite, we have $\rqnorm{R(\rho)}_1 = \mathrm{tr}\left(\mathcal{R}(
\rho
)\right)$, which allows us to express the above inequality as:
\begin{equation}
    \begin{split}
        &\left|\frac{1}{q}-\frac{1}{p}\right|q + \left\|\frac{\mathcal{R}(\rho)}{p} - \frac{\mathcal{U}(\rho)}{p}\right\|_1\\
        &=\frac{\left|\mathrm{tr}\left((\mathcal{R}-\mathcal{U})(\rho)\right)\right|}{p} + \left\|\frac{\mathcal{R}(\rho)}{p} - \frac{\mathcal{U}(\rho)}{p}\right\|_1\\
        &\leq \frac{2}{p}\left\|\left(\mathcal{R}-\mathcal{U}\right)(\rho)\right\|_1.
    \end{split}
\end{equation}Using the definition of the diamond norm as a supremum over the trace norm of the map (with an auxillary identity map), we have:
\begin{equation}
    \begin{split}
        &\left\|\frac{\mathcal{R}(\rho)}{q}-\frac{\mathcal{U}(\rho)}{p}\right\|_1\leq \frac{2}{p}\left\|\mathcal{R} - \mathcal{U}\right\|_\diamond\leq\frac{2\epsilon}{p}.
    \end{split}
\end{equation}
\end{proof}The application to the settings in Thm.~\ref{rq:thm:main} and Thm.~\ref{thm: DR-QSVT} (separately) is immediate. That is, if we have a diamond norm error bound $\epsilon$ for the un-normalized maps, then to predict the maximum trace distance between two output states corresponding to the randomized and deterministic QSVT maps, we have apply a factor of $p^{-1}$. We note that this also applies to the polynomial sampling method of Ref.~\cite{martyn_halving_2025}. 
\subsection{Error bound prefactors for particular function families}

The error bounds of RBE-QSVT and DR-QSVT both contain a polynomial-dependent prefactor defined in Eq.~\eqref{eq: polynomial prefactor}. The larger this prefactor is, the smaller the ratio $(\lambda_\epsilon/\lambda)^2/n_T$ must be in order to obtain performance guarantees for the method, which corresponds to the necessity of including more terms of the block-encoded operator for the QSVT circuits, which often corresponds to deeper circuits. This prefactor depends on both the chosen polynomial $P$ and the extension $g$ of $P^\prime$ over $x\in\mathbb{R}$. For families of functions for which the extension $g$ is just trivially $P^\prime$ itself over the entire real line, we can easily study how $C_P$ scales with respect to the "difficulty" of implementing $P$ in a QSVT circuit. For example, for Hamiltonian dynamics one can separately prepare $\mathrm{cos}(xt)$ and $\mathrm{sin}(xt)$, then take a linear combination of QSVT circuits to obtain $e^{ixt} = \mathrm{cos}(xt)+i\mathrm{sin}(xt)$ through Euler's identity. The QSP degree one needs is characterized by the evolution time $t$. Similarly, one could prepare the functions $\mathrm{erf}(kx)$ or $e^{-\beta x^2}$ for filtering problems such as ground state preparation. In these instances, a larger $k$ or $\beta$ requires a larger QSP degree.

In Appendix~\ref{sec: prefactor calculations}, we consider the exact target function $f$ and explicitly compute $C_f$ for the functions $\mathrm{cos}(xt)$, $\mathrm{erf}(kx)$, and $e^{-\beta x^2}$. These prefactors are:
\begin{equation}\label{eq: function c_p scaling}
\begin{aligned}
    C_{\mathrm{cos},t} &= 2t^2\\
    C_{\mathrm{erf},k} &=\frac{4}{\pi}\left(1+\mathrm{erf}(k)\right)k^2\\
    C_{\mathrm{Gauss},\beta} &=\left(\frac{4}{\pi}+2\right)\beta.
\end{aligned}
\end{equation}We note that these prefactors are for common target functions, not finite polynomial approximations. In practice, given a finite polynomial approximation of a target function, we would numerically compute $C_P$ using an extension function such as the one given in Appendix~\ref{sec: g construction}. We note that these functions do not obey the assumptions of Thm.~\ref{rq:thm:main} as they do not have compact support. Nonetheless, these expressions still have utility in understanding the relative difficulty in implementing randomized QSVT implementations of various target functions.

\section{Applications to Lattice Models and Quantum Chemistry}
\label{sec:numerical}

In Sec.~\ref{sec: rand qsvt method} we derived performance guarantees for RBE-QSVT and DR-QSVT in the form of error bounds that can be efficiently computed using only information about the choice of QSVT polynomial and the coefficients of the operator to be block-encoded. Importantly, computing these bounds does not require any knowledge about the singular values of the block-encoded operator. In this section, we numerically compute the bounds for RBE-QSVT as a fraction of the number of Pauli terms in random block-encoded Hamiltonians. \emph{i.e.} we have a Hamiltonian $H$ with $L$ terms, select a core Hamiltonian with $L_0$ terms, then sample $n_T$ Pauli operators (with replacement) from the tail $H_\epsilon$. We are primarily interested in how the error bound in Thm.~\ref{rq:thm:main} scales as a function of $(L_0 + n_T)/L$. We emphasize that the numerical calculations shown here denote proven performance guarantees in the form of error bounds, not the exact errors computed through explicit simulation of the method, which could be much lower than the error bounds. The error bounds we have proven are not guaranteed to be tight.

\subsection{Methods}

We study two classes of Hamiltonians in this section:  \textbf{1.} Heisenberg Hamiltonians with long-range interactions on an $8\times 8$ square grid with open boundary conditions and \textbf{2.} electronic structure Hamiltonians of iron-sulfur complexes. We include the former to investigate which types of long-range interactions governed by various power-laws may be particularly benefited by randomization. We include the latter as they constitute a class of well-known strongly-correlated many-electron systems which have long been studied as targets for quantum computers~\cite{robledo-moreno_chemistry_2025, Motta2021, Lee2023, reiher_elucidating_2017, Low_fast_quantum_simulation, kanasugi2026enabling, wecker2014gate}.
In all numerical calculations, for a fixed budget $B$ we choose the $(L_0, n_T)$ such that $L_0 + n_T = B$ which minimizes the error bound in Thm.~\ref{rq:thm:main}. The particular extension $g$ of $P^\prime$ as described in Thm.~\ref{rq:thm:main} is that given in Appendix~\ref{sec: g construction} with width $w$ optimized to minimize the error bound. The exact widths $w$ used in this work can be found in Table~\ref{tab:polynomial-degrees-widths}. We limit the study to RBE-QSVT for the sake of conciseness. A qualitative picture of the performance guarantees of DR-QSVT can be inferred by combining the RBE-QSVT results in this work with the numerical results studying random polynomial sampling in Ref.~\cite{martyn_halving_2025}.

We study four classes of target functions: $\mathrm{cos}(kx)$, $e^{-\beta}\mathrm{cosh}(\beta x)$, $\mathrm{erf}(kx)$, and $e^{-\beta x^2}$. The results for the error and hyperbolic cosine functions are deferred to Appendix~\ref{sec: additional error bound plots} as they are qualitatively similar to those of $\mathrm{cos}(kx)$. These are chosen for their relevance to real-time Hamiltonian dynamics, imaginary-time evolution, and filtering problems. The finite polynomial approximation degrees of these functions is chosen such that the approximation error is less than $10^{-4}$. That is, we sweep degrees $2^j$ ($2^j-1$ for odd functions) and search for the smallest $j$ such that the approximation error is less than $10^{-4}$. We then sweep $j^\prime = j, j+1, j+2$ for the value $j^\prime$ which minimizes $C_P$. The reason for this is that our choice of extension $g$ is sensitive to $P^\prime(\pm 1)$. Thus, choosing a degree which approximates the target function to small error can help stabilize the estimate of $C_P$. Trailing Chebyshev terms with coefficients of exactly zero are removed, which can reduce the degree in some edge cases. The exact polynomial degrees used for all target functions in this work can be found in Table~\ref{tab:polynomial-degrees-widths}. Throughout this section, we use $\delta$ as shorthand for $\lambda_\epsilon/\lambda$. Hamiltonians are represented using Pennylane 0.45.1~\cite{bergholm2018pennylane}. For each Hamiltonian, we truncate operator terms with coefficient magnitudes below $10^{-12}$.

\begin{table}[htbp]
\centering
\small
\begin{tabular}{l|r|r|l}
\hline
$f(x)$ & Parameter $k/\beta$ & Degree $d$ & Width $w$ \\
\hline
$\cos(kx)$
 & 4    & 32   & $7.46 \times 10^{-1}$ \\
 & 16   & 128  & $2.09 \times 10^{-1}$ \\
 & 64   & 128  & $4.70 \times 10^{-2}$ \\
 & 128  & 256  & $5.14 \times 10^{-2}$ \\
 & 1024 & 2048 & $4.00 \times 10^{-3}$ \\
\hline
$\mathrm{erf}(kx)$
 & 4    & 127  & $2.74 \times 10^{-3}$ \\
 & 16   & 255  & $3.34 \times 10^{-4}$ \\
 & 64   & 511  & $2.41 \times 10^{-4}$ \\
 & 128  & 2047 & $1.60 \times 10^{-5}$ \\
 & 1024 & 8191 & $1.78 \times 10^{-5}$ \\
\hline
$e^{-\beta}\cosh(\beta x)$
 & 4    & 8    & $5.20 \times 10^{-1}$ \\
 & 16   & 16   & $1.32 \times 10^{-1}$ \\
 & 64   & 32   & $3.34 \times 10^{-2}$ \\
 & 128  & 64   & $1.66 \times 10^{-2}$ \\
 & 1024 & 128  & $2.11 \times 10^{-3}$ \\
\hline
$e^{-\beta x^2}$
 & 4    & 32   & $4.09 \times 10^{-1}$ \\
 & 16   & 128  & $6.04 \times 10^{-3}$ \\
 & 64   & 128  & $2.21 \times 10^{-5}$ \\
 & 128  & 256  & $3.11 \times 10^{-5}$ \\
 & 1024 & 510  & $2.13 \times 10^{-5}$ \\
\hline
\end{tabular}
\caption{Finite polynomial approximation degrees and widths $w$ used for the extension function $g$ used throughout the numerical calculations.}
\label{tab:polynomial-degrees-widths}
\end{table}

\subsection{Long-Range Heisenberg Model}
The Hamiltonian studied in this section is given by:
\begin{equation}\label{eq: long-range heisenberg}
    H_{\alpha} = \sum_{i< j}\frac{1}{r_{ij}^\alpha}\left(X_iX_j+Y_iY_j+Z_iZ_j\right)
\end{equation}where $r_{ij}$ is the Cartesian distance between lattice points $i$ and $j$ in units of the length along an edge in the square lattice. \emph{e.g.} $r_{ij}=1$ for two adjacent lattice sites in a square lattice and $r_{ij}={\sqrt{2}}$ for $i,j$ on opposite sides of the diagonal bisecting a square unit cell. $\alpha \in \mathbb{R}$ denotes the power-law of the long-range interactions. In this section we consider an $8\times 8$ two-dimensional square lattice.

\begin{figure*}[ht!]\includegraphics[width=0.75\linewidth]{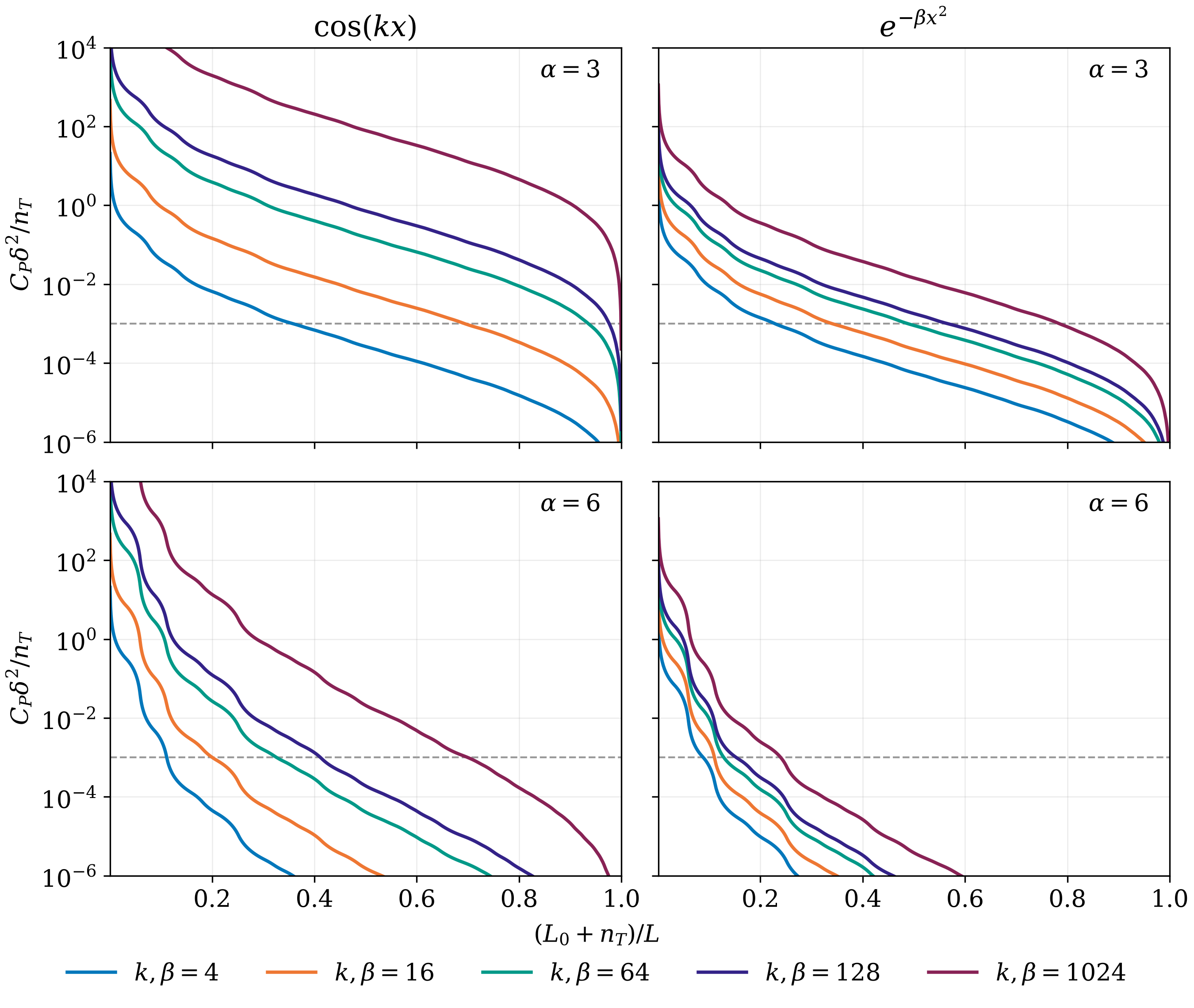}
    \caption{The diamond norm error bound in Thm.~\ref{rq:thm:main} as a function of ($L_0+n_T$)/L for the long-range Heisenberg model in Eq.~\eqref{eq: long-range heisenberg} for the cosine and Gaussian filter functions. Here, $\delta = \lambda_\epsilon/\lambda$. The value of $\alpha$ governs the power-law of the long-range interactions. The horizontal grey dashed line denotes a target error bound of $10^{-3}$. The left and right columns correspond to the cosine and Gaussian filtering functions, respectively.}
    \label{fig:long_range_heisenberg_error_scaling}
\end{figure*}

The results for the long-range Heisenberg model with $\alpha = 3,6$ are shown in Fig.~\ref{fig:long_range_heisenberg_error_scaling}.
We can see that there is a clear trade-off between the "difficulty" of a given target function and the number of Pauli terms one needs to include in the random block-encoded Hamiltonian. For example, for $\mathrm{cos}(kx)$, larger $k$ corresponds to longer evolution times for real-time Hamiltonian dynamics. We can see that for $\alpha = 6$ and a target error of $10^{-3}$, $k=4$ requires retaining only $\sim 0.1$ of the Pauli terms, whereas $k=128$ and $k=1024$ require retaining $\sim 0.4$ and $\sim 0.7$ of the Pauli terms, respectively. Similar trends persists for the Gaussian filter. These trends are intuitive when viewed through the lens of Eq.~\eqref{eq: function c_p scaling}, which shows that the polynomial-dependent prefactor $C_P$ scales as $\mathcal{O}(k^2)$ for $\mathrm{cos}(kx)$ and $\mathcal{O}(\beta)$ for $e^{-\beta x^2}$. The larger $C_p$ is, the smaller $\delta^2/n_T$ needs to be in order to achieve a given target error, which typically requires more retained Pauli terms. As the Gaussian filter is the only function considered here with $C_p$ scaling linearly with its difficulty parameter $\beta$, we need to retain fewer terms for a fixed Hamiltonian and target function. Whereas $\mathrm{cos}(kx)$ needs to retain $\sim 0.7$ of the Pauli terms for $k=1024$ for a target error of $10^{-3}$, the Gaussian filter only requires $\sim 0.25$ of the terms for $\alpha =6$.

\begin{figure}[ht!]
    \centering
    \includegraphics[width=0.95\linewidth]{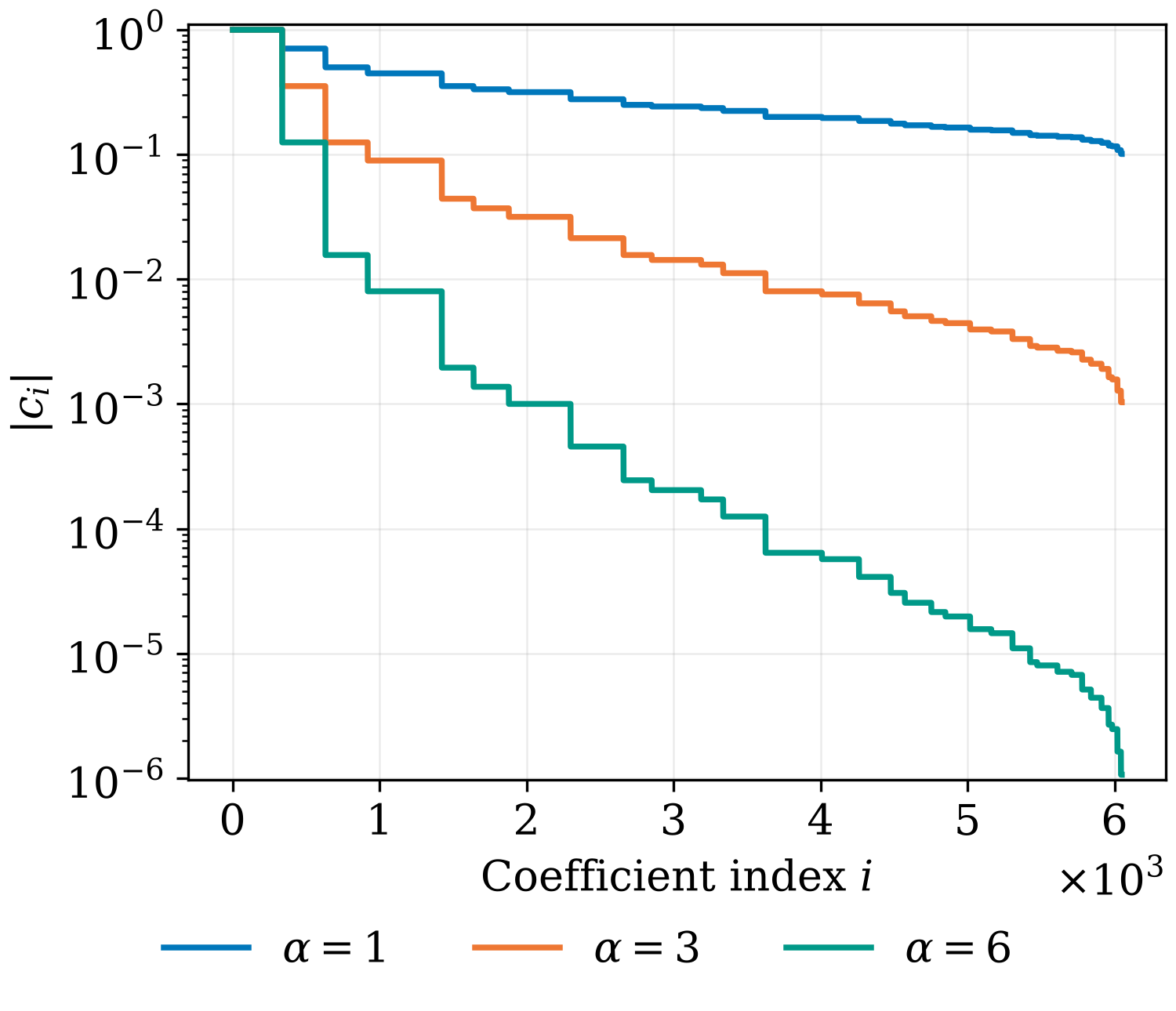}
    \caption{The magnitude of the coefficients of the long-range Heisenberg model Hamiltonians for $\alpha=1,3,6$, ordered in terms of descending magnitude.}
    \label{fig: heisenberg coefficient scaling}
\end{figure}

We also comment that the power-law scaling of the long-range interactions is decisive with regards to which Hamiltonians we can prove will benefit most from RBE-QSVT. Note that for $\alpha = 3$ and a target error of $10^{-3}$, a larger number of Pauli terms is typically required compared to $\alpha = 6$, regardless of the target function. These trends are intuitive in the sense that if interaction terms are of the form $V(r)\sim 1/r^\alpha$, then a larger $\alpha$ naturally gives rise to a smaller number of Hamiltonian terms with large weight and a larger number of terms which decay more quickly with the distance between lattice sites. We quantify this intuition in Fig.~\ref{fig: heisenberg coefficient scaling}, which shows more rapid decay of the magnitude of the coefficients for larger $\alpha$, as expected. For example, the magnitude of the coefficients for $\alpha=1$ never drop below $\sim 0.1$, whereas those of $\alpha=3,6$ span the ranges of $\sim 10^{-3}-10^0$ and $\sim 10^{-6}-10^0$, respectively.

\begin{figure*}[phtb!]
    \centering
    \includegraphics[width=0.75\linewidth]{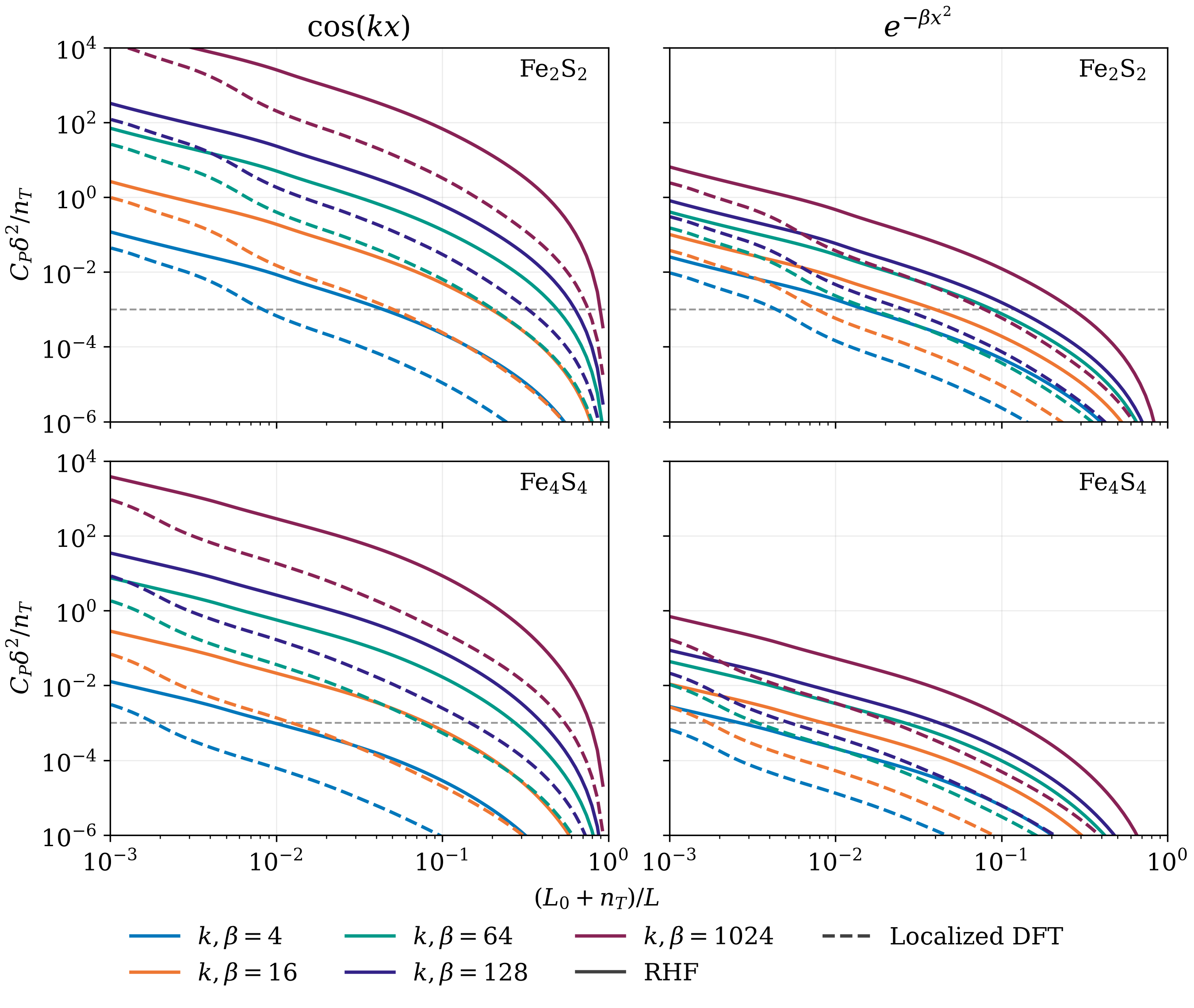}
    \caption{The diamond norm error bounds from Thm.~\ref{rq:thm:main} as a function of the fractional Pauli term budget for the iron-sulfur complexes $\mathrm{Fe_2S_2}$ and $\mathrm{Fe_4S_4}$ for the cosine and Gaussian filtering functions. Dashed lines denote the localized DFT orbitals described in Ref.~\cite{li_chan_fcidump} and solid lines denote the transformed MO orbitals from Ref.~\cite{robledo-moreno_chemistry_2025} in the same active space. The left and right columns correspond to the cosine and Gaussian filtering functions, respectively.}
    \label{fig:iron-sulfur error plots}
\end{figure*}

\subsection{Iron-Sulfur Complexes}\label{sec: iron-sulfur numerical calculations}

Based on the results of the $\alpha = 3$ long-range Heisenberg model in the previous section, one might be tempted to think that the whole of electronic structure Hamiltonians, which have Coulomb interactions of the form $V(r)\sim 1/r$, likely could not meaningfully benefit from randomized block-encodings. In this section, we show numerically that this is not the case and that demanding that interactions decay as $1/r^\alpha$ for $\alpha \geq 3$ is in general excessively stringent. We further show that the degree to which randomized block-encodings are advantageous depends to a large extent on the choice of basis set.

We calculate the error bounds for the iron-sulfur complexes $\mathrm{[Fe_2S_2(SCH_3)_4]^{2-}}$ (20o, 30e) and $\mathrm{[Fe_4S_4(SCH_3)_4]^{2-}}$ (36o, 54e) for Hamiltonians generated from the publicly available FCIDUMP files of Ref.~\cite{robledo-moreno_chemistry_2025} and Ref.~\cite{li_chan_fcidump}. Ref.~\cite{robledo-moreno_chemistry_2025} uses the active space integrals derived from a localized DFT calculation in Ref.~\cite{li_chan_fcidump} and transforms them to the canonical basis using a restricted Hartree-Fock calculation. Thus, including both of these orbital sets in the current study is a convenient means of investigating the choice of basis for a fixed active space. Hamiltonians are constructed using only the electronic contributions to the energy and do not include the constant nuclear repulsion energy term.

\begin{figure}[htpb!]
    \centering
    \includegraphics[width=0.95\linewidth]{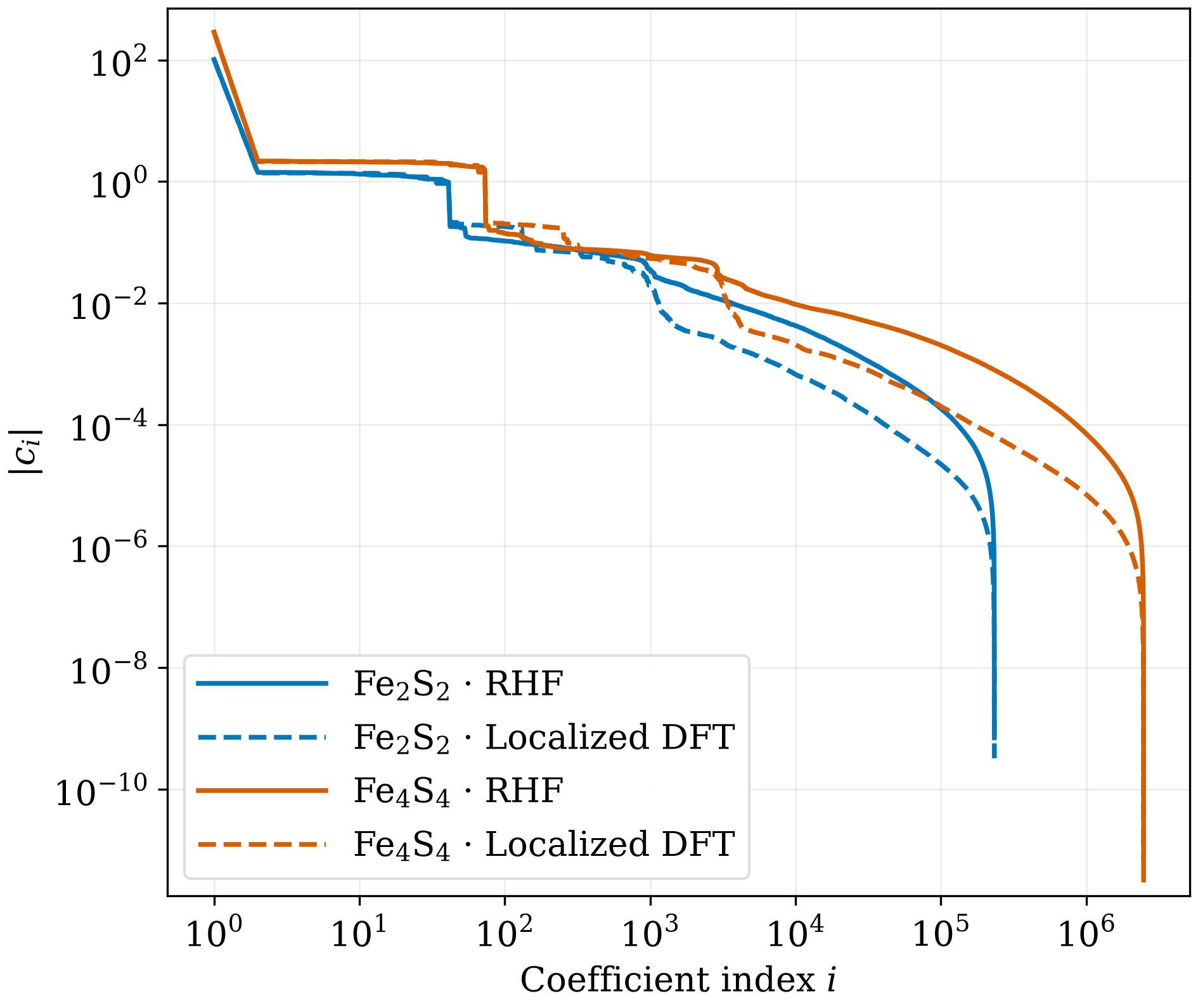}
    \caption{The magnitude of the coefficients of the Hamiltonians of the iron-sulfur complexes for $\alpha=1,3,6$, ordered in terms of descending magnitude. The solid blue and orange dashed lines denote Hamiltonians represented in the orbital basis sets of Ref.~\cite{robledo-moreno_chemistry_2025} and Ref.~\cite{li_chan_fcidump}, respectively.}
    \label{fig:iron-sulfur coefficients plot}
\end{figure}

We can see from Fig.~\ref{fig:iron-sulfur error plots} that using the localized DFT orbitals of Ref.~\cite{li_chan_fcidump} is advantageous as compared to using RHF orbitals in essentially all cases studied. For example, we can see that for the Gaussian filter function with $\beta = 1024$, the target error of $10^{-3}$ is achieved with a fractional Pauli term budget of $\sim 0.02$, whereas RHF orbitals require a fractional budget of $\sim 0.1$. Similar statements can be made to varying degrees for the other target functions studied. This effect can be understood from the context given by Fig.~\ref{fig:iron-sulfur coefficients plot}, which shows that both orbital sets obey similar coefficient power-law scaling, but to differ by a constant factor of $\sim5-10$ after coefficient indices $\sim 10^3 - 10^4$. This constant factor is evidently sufficient to give rise to a meaningful advantage in the context of RBE-QSVT. 

These plots illustrate performance guarantees with respect to how sparse the block-encoded operator can be made as a function of the target error. The extent to which this improved sparsity translates into circuit depth and runtime reductions depends on the particular block-encoding method used. For example, using an LCU-style block-encoding, the circuit depth would be reduced by a $\mathcal{O}((L_0+n_T)/L)$ multiplicative factor.

\section{Discussion and Conclusions}\label{sec: conclusions}

In this work we have proposed two variants of randomized QSVT which utilize randomized block-encodings of linear operators to reduce the cost of block-encoding for early fault-tolerant quantum computers. The first method, RBE-QSVT block-encodes an operator comprised of the sum of a core operator $A_0$ and a term proportional to the average of $n_T$ randomly sampled terms from a tail operator $A_\epsilon$. Thus, under the assumption that one uses a block-encoding whose circuit depth is dependent on the number of terms in the operator (such as LCU where the complexity scales with the number of terms $L$ as $\mathcal{O}(L)$), one can potentially reduce the circuit depth needed to perform QSVT considerably. The second, DR-QSVT, samples QSVT circuits sampled from a joint distribution on both block-encoded operators and polynomials. This allows one to inherit the circuit depth savings of both having fewer queries to the block-encoding as well as a less expensive block-encoding itself. This comes at the cost of only additive error contributions from each type of randomization.

We have further proven rigorous error bounds for both methods which provide performance guarantees on a polynomial by polynomial basis. That is, given a polynomial $P$ and target error $\epsilon$, one can calculate the smallest number of core terms $L_0$ such that $\tfrac{C_P}{n_T}\left(\frac{\lambda_\epsilon}{\lambda}\right)^2 \leq \epsilon$. One can similarly compute an optimal $L_0$ and average polynomial degree $d_{avg}$ for DR-QSVT.

We have numerically computed error bounds for long-range Heisenberg Hamiltonians and the electronic structure Hamiltonians for iron-sulfur complexes. We have found that in several cases, the number of Pauli terms that need to be retained can be reduced by an order of magnitude or more, depending on the Hamiltonian and target function used. Thus, we reasonably expect this to correspond to a significant reduction in the circuit depth needed to simulate these systems on early fault-tolerant quantum computers.

There are several directions one could take this work in the future. The first would be to investigate how tight the bounds we have derived are and improve upon them if doing so would be well-motivated. We reiterate that the bounds derived here are not guaranteed to be tight and that improving the error bounds by even one to two orders of magnitude would in several cases allow one to prove that significantly fewer Pauli terms could be retained for a fixed target error bound. The challenge in numerically testing the tightness of these bounds is that known tighter bounds such as the mixing lemma or the exact diamond norm error require computing quantities such as the full eigenspectrum of the Hamiltonian of interest. This limits numerical studies to relatively modestly-sized systems. It is unclear how much information about utility-scale systems ($\gtrapprox 50$ qubits) one could glean from this information.

The second direction would be to investigate optimizing the basis in which the Hamiltonian is represented to maximize the reduction in circuit depth offered by RBE-QSVT. For example, we saw in Sec.~\ref{sec: iron-sulfur numerical calculations} that simply changing the orbital basis from the canonical orbitals of Ref.~\cite{robledo-moreno_chemistry_2025} to those derived from localized DFT in Ref.~\cite{li_chan_fcidump} allowed one to significantly reduce the number of retained Pauli terms. This improvement is incidental rather than deliberate. Constructing change-of-basis algorithms that are deliberately optimized for randomized block-encodings could allow one to see significant further gains from RBE-QSVT and similar methods.
Finally, as the methods presented in this work open has the potential to greatly reduce the circuit depth of QSVT, it would be interesting to experimentally realize this method on current-day hardware. We look forward to future advances along this direction of research.

\begin{acknowledgements}
    This work utilized large language models (LLMs) in the derivation of the error bounds for the two methods in this work, generating code for the numerical calculations, transcribing handwritten notes to LaTeX code, hypothesis testing and test system search, checking for errors and reproducibility of results, and general manuscript revision suggestions. This work is supported by the U.S. Department of Energy, Office of Science, Advanced Scientific Computing Research, under contract number DE-SC0025384.
\end{acknowledgements}

\bibliography{ref}

\appendix

\section{Helper Lemmas for Main Results}\label{sec: helper lemmas}

\subsection{Mixing bound}

In the following lemma, we parallel the mixing lemma proofs of
Hastings~\cite{hastings_turning_2016}, Campbell~\cite{campbell_shorter_2017}, and Martyn-Rall~\cite{martyn_halving_2025} for our particular case.

\begin{lemma}[Mixing lemma for $p_\ell$ and $P(A_\ell)$]
\label{rq:lem:mixing}
With the set-up in Sec.~\ref{sec: main method} and definitions of Eq.~\eqref{rq:eq:errors}, we can bound the diamond norm error by
\begin{equation}
 \rqnorm{\mathcal{R}_P-\mathcal{U}_P}_\diamond\leq q_P+2\rqnorm{P^{\sv}(\rqA)}b_P.
\label{rq:eq:mixing}
\end{equation}
\end{lemma}

\begin{proof}
Write $P^{\sv}(A_\ell)=P^{\sv}(\rqA)+\Delta_\ell$ and expand:
\begin{equation}
\begin{aligned}
 &\rqE\!\left[P^{\sv}(A_\ell)\rho P^{\sv}(A_\ell)^\dagger\right]\\
 &\quad=\rqE\!\left[(P^{\sv}(\rqA)+\Delta_\ell)
       \rho(P^{\sv}(\rqA)+\Delta_\ell)^\dagger\right]\\
 &\quad=P^{\sv}(\rqA)\rho P^{\sv}(\rqA)^\dagger
       +\rqE[\Delta_\ell]\rho P^{\sv}(\rqA)^\dagger\\
 &\qquad\quad+P^{\sv}(\rqA)\rho\rqE[\Delta_\ell^\dagger]
       +\rqE[\Delta_\ell\rho\Delta_\ell^\dagger].
\end{aligned}
\label{rq:eq:mixing-expansion}
\end{equation}
Thus
\begin{equation}
\begin{aligned}
 (\mathcal{R}_P-\mathcal{U}_P)(\rho)
 &=\rqDelta\rho P^{\sv}(\rqA)^\dagger
   +P^{\sv}(\rqA)\rho\rqDelta^\dagger\\
 &\quad+\rqE[\Delta_\ell\rho\Delta_\ell^\dagger].
\end{aligned}
\label{rq:eq:map-difference}
\end{equation}
Since $\Phi(\rho)=\rqE[\Delta_\ell\rho\Delta_\ell^\dagger]$ is
completely positive, by [\cite{watrous2018theory}~Definition 3.43 and Theorem 3.39], we have
\begin{equation}
\begin{aligned}
 \rqnorm{\Phi}_\diamond
 &=\rqnorm{\Phi^\dagger(\rqI)}\\
 &=\rqnorm*{\rqE[\Delta_\ell^\dagger\Delta_\ell]}=q_P.
\end{aligned}
\label{rq:eq:cp-norm}
\end{equation}We further have
\begin{equation}
\begin{aligned}
 \rqnorm{\rqDelta\rho P^{\sv}(\rqA)^\dagger}_1
 &,\rqnorm{P^{\sv}(\rqA)\rho\rqDelta^\dagger}_1\\
 &\leq\rqnorm{\rqDelta}\rqnorm{\rho}_1\rqnorm{P^{\sv}(\rqA)}
\end{aligned}
\label{rq:eq:cross-term}
\end{equation}by [\cite{watrous2018theory},~Eq.~1.175]. Thus, defining $\mathcal{E}(\rho)=\rqDelta\rho P^{\sv}(\rqA)^\dagger$,
we have
\begin{equation}
\begin{aligned}
 \rqnorm{\mathcal{E}}_\diamond
 &=\sup_\rho\frac{\rqnorm{(\mathcal{E}\otimes\mathcal{I})(\rho)}_1}{\left\|\rho\right\|_1}\\
 &\leq\rqnorm{\rqDelta}\rqnorm{P^{\sv}(\rqA)}\\
 &=\rqnorm{P^{\sv}(\rqA)}b_P.
\end{aligned}
\label{rq:eq:cross-diamond}
\end{equation}where we have used H\"older's inequality. Consequently,
\begin{equation}
 \rqnorm{\mathcal{R}_P-\mathcal{U}_P}_\diamond
 \leq q_P+2\rqnorm{P^{\sv}(\rqA)}b_P.
 \label{rq:eq:mixing-conclusion}
 \qedhere
\end{equation}
\end{proof}

\subsection{Difference of exponentials}\label{sec: diff of exp}
\emph{Setting.} Consider an ensemble $\{p_\ell, \tilde{H}_\ell\}$ on hermitian operators $\tilde{H}_\ell$ such that $\mathbb{E}\left[\tilde{H}_\ell\right] = \tilde{\bar{H}}$ and $\rqnorm{\tilde{H}_\ell}\leq 1~\forall~\ell$. Take $\tilde{\Delta}_\ell = P(\tilde{H}_\ell) - P(\tilde{\bar{H}})$, $\tilde{E}_\ell = \tilde{H}_\ell - \tilde{\bar{H}}$, $\tilde{V} = \rqE[\tilde{E}_\ell^\dagger\tilde{E}_\ell]$,  $\tilde{\nu} = \rqnorm{\tilde{V}}$, $\bar{\tilde{\Delta}} = \mathbb{E}[\tilde{\Delta}_\ell]$,  $\tilde{q}_P=\rqnorm{\rqE[\tilde{\Delta}_\ell^\dagger\tilde{\Delta}_\ell]}$, and $\tilde{b}_P=\rqnorm{\mathbb{E}[\tilde{\Delta}_\ell]}$.
\begin{lemma}
\label{rq:lem:duhamel}
Consider the above setting. For any $t\in\mathbb{R}$,
\begin{equation}
 e^{it\tilde{H}_\ell}-e^{it\tilde{\bar{H}}}
 =it\int_0^1e^{it(1-u)\tilde{H}_\ell}\tilde{E}_\ell e^{itu\tilde{\bar{H}}}\,du.
\label{rq:eq:duhamel}
\end{equation}
\end{lemma}

\begin{proof}
We note the identity in operator function theory from Ref.~\cite{Lipschitz}:
\begin{equation}
    e^{iAt} - e^{iBt} = i\int_0^t e^{iA(t-s)}(A-B)e^{iBs}
ds\end{equation}
for operators $A$ and $B$. This identity can be understood as a specialization of the variation of parameters formula in Lemma 1 of Ref.~\cite{Childs_2021}. Identifying $A$ with $\tilde{H}_\ell$, $B$ with $\tilde{\bar{H}}$, and $A-B$ with $\tilde{E}_\ell$ gives us
\begin{equation}
e^{it\tilde{H}_\ell} - e^{it\tilde{\bar{H}}} = i\int_0^t e^{i(t-s)\tilde{H}_\ell}\tilde{E}_\ell e^{is\tilde{\bar{H}}}ds.
\end{equation}The change of parameters $s = tu$ in this integral gives Eq.~\eqref{rq:eq:duhamel}.
\end{proof}

\subsection{Bounds on magnitude of $\rqE[\tilde{E}_\ell Z_\ell \tilde{E}_\ell]$}
\begin{lemma}
\label{rq:lem:block}
Consider the setting of Sec.~\ref{sec: diff of exp}. Define the block column
\begin{equation}
 S_E=
 \begin{pmatrix}
  \sqrt{p_1}\tilde{E}_1\\
  \sqrt{p_2}\tilde{E}_2\\
  \vdots
 \end{pmatrix}.
\label{rq:eq:SE}
\end{equation}
Thus $(S_E)_\ell=\sqrt{p_\ell}\tilde{E}_\ell$ and
\begin{equation}
 S_E^\dagger S_E
 =\rqE[\tilde{E}_\ell^2]=\tilde{V}.
\label{rq:eq:SE-gram}
\end{equation}
For arbitrary randomly sampled operators $Z_\ell$,
\begin{equation}
 \rqnorm*{\rqE[\tilde{E}_\ell Z_\ell \tilde{E}_\ell]}
 \leq\tilde{\nu}\max_\ell\rqnorm{Z_\ell}.
\label{rq:eq:block-bound}
\end{equation}
\end{lemma}
\begin{proof}
Let $\rqdiag(Z_\ell)$ denote the block-diagonal matrix
\begin{equation}
 \rqdiag(Z_\ell)=
 \begin{pmatrix}
  Z_1&0&\cdots\\
  0&Z_2&\ddots\\
  \vdots&\ddots&\ddots
 \end{pmatrix}.
\label{rq:eq:block-diagonal}
\end{equation}
Then
\begin{equation}
 \rqE[\tilde{E}_\ell Z_\ell \tilde{E}_\ell]
 =S_E^\dagger\rqdiag(Z_\ell)S_E.
\label{rq:eq:block-factorization}
\end{equation}
Thus
\begin{equation}
\begin{aligned}
 \rqnorm*{\rqE[\tilde{E}_\ell Z_\ell \tilde{E}_\ell]}
 &=\rqnorm{S_E^\dagger\rqdiag(Z_\ell)S_E}\\
 &\leq\rqnorm{S_E}^2\rqnorm{\rqdiag(Z_\ell)}\\
 &\leq\tilde{\nu}\max_\ell\rqnorm{Z_\ell}.
\end{aligned}
\label{rq:eq:block-proof}
\end{equation}
\end{proof}

\subsection{Bound on \texorpdfstring{$\tilde{q}_P$}{qP}}
\begin{lemma}
\label{rq:lem:variance}
Consider the setting of Sec.~\ref{sec: diff of exp}. Take $g$ as in Theorem.~\ref{rq:thm:main}. Then
\begin{equation}
 \tilde{q}_P\leq\tilde{\frac{\nu}{2\pi}} \rqnorm{\widehat{g}}_1^2.
\label{rq:eq:qP}
\end{equation}
\end{lemma}

\begin{proof}
Let
\begin{equation}
 P(x)=P(0)+\int_0^xg(s)\,ds,\qquad x\in[-1,1].
\label{rq:eq:primitive}
\end{equation}
Taking the inverse Fourier transform of $g$ gives
\begin{equation}
\begin{aligned}
 P(x)
 &=P(0)+\frac{1}{\sqrt{2\pi}}\int_{\mathbb{R}}\widehat{g}(t)
       \left(\int_0^x e^{its}\,ds\right)dt\\
 &=P(0)+\frac{1}{\sqrt{2\pi}}\int_{\mathbb{R}}
       \widehat{g}(t)\frac{e^{itx}-1}{it}\,dt.
\end{aligned}
\label{rq:eq:scalar-functional-calculus}
\end{equation}
For a Hermitian matrix $X$ with eigenspectrum contained in $[-1,1]$, this yields
\begin{equation}
 P(X)=P(0)\rqI+\frac{1}{\sqrt{2\pi}}\int_{\mathbb{R}}
       \widehat{g}(t)\frac{e^{itX}-\rqI}{it}\,dt.
\label{rq:eq:matrix-functional-calculus}
\end{equation}
We refer the reader to Ref.~\cite{Kissin_Shulman_2005} and Ref.~\cite{aleksandrov2016operator} for background on bounding the difference between operator functions using norms of Fourier transforms and the machinery of operator Lipschitz functions more generally. Thus
\begin{equation}
\begin{aligned}
 &P(\tilde{H}_\ell)-P(\tilde{\bar{H}})\\
 &\quad=\frac{1}{\sqrt{2\pi}}\int_{\mathbb{R}}
       \frac{\widehat{g}(t)}{it}
       \left(e^{it\tilde{H}_\ell}-e^{it\tilde{\bar{H}}}\right)dt.
\end{aligned}
\label{rq:eq:delta-fourier}
\end{equation}
By Lemma~\ref{rq:lem:duhamel},
\begin{equation}
\begin{aligned}
 \tilde{\Delta}_\ell&=P(\tilde{H}_\ell)-P(\tilde{\bar{H}})\\
 &=\frac{1}{\sqrt{2\pi}}\int_{\mathbb{R}}\widehat{g}(t)
   \int_0^1 e^{it(1-u)\tilde{H}_\ell}\tilde{E}_\ell e^{itu\tilde{\bar{H}}}\,du\,dt.
\end{aligned}
\label{rq:eq:delta-integral}
\end{equation}
Define
\begin{equation}
 S_{\tilde{\Delta}}=
 \begin{pmatrix}
  \sqrt{p_1}\tilde{\Delta}_1\\
  \sqrt{p_2}\tilde{\Delta}_2\\
  \vdots
 \end{pmatrix},
 \qquad\rqnorm{S_{\tilde{\Delta}}}^2=\tilde{q}_P.
\label{rq:eq:Sdelta}
\end{equation}
Define the block-diagonal operator
\begin{equation}
 D(t,u)=\rqdiag_\ell\!\left(e^{it(1-u)\tilde{H}_\ell}\right).
\label{rq:eq:block-unitary}
\end{equation}\emph{i.e.} $D(t,u)$ is the block-diagonal matrix whose $\ell$th block is $e^{it(1-u)\tilde{H}_\ell}$. Then
\begin{equation}
 S_{\tilde{\Delta}}=\frac{1}{\sqrt{2\pi}}\int_{\mathbb{R}}\widehat{g}(t)
   \int_0^1D(t,u)S_E e^{itu\tilde{\bar{H}}}\,du\,dt,
\label{rq:eq:Sdelta-integral}
\end{equation}
and hence
\begin{equation}
\begin{aligned}
 \rqnorm{S_{\tilde{\Delta}}}
 &\leq\frac{1}{\sqrt{2\pi}}\int_{\mathbb{R}}\rqabs{\widehat{g}(t)}\\
 &\qquad\times\int_0^1
   \rqnorm{D(t,u)S_E e^{itu\tilde{\bar{H}}}}\,du\,dt.
\end{aligned}
\label{rq:eq:Sdelta-bound}
\end{equation}
We have:
\begin{equation}
\begin{aligned}
 \rqnorm{D(t,u)S_E e^{itu\tilde{\bar{H}}}}
 &\leq\rqnorm{S_E}\\
 &=\sqrt{\rqnorm{\tilde{V}}}=\sqrt{\tilde{\nu}}.
\end{aligned}
\label{rq:eq:block-integrand-bound}
\end{equation}
Therefore
\begin{equation}
 \rqnorm{S_{\tilde{\Delta}}}
 \leq\frac{\sqrt{\tilde{\nu}}}{\sqrt{2\pi}}\int_{\mathbb{R}}
      \rqabs{\widehat{g}(t)}\,dt
 =\sqrt{\frac{\tilde{\nu}}{2\pi}}\,\rqnorm{\widehat{g}}_1.
\label{rq:eq:Sdelta-Ag}
\end{equation}
\end{proof}

\subsection{Bound on \texorpdfstring{$\tilde{b}_P$}{bP}}
\begin{lemma}
\label{rq:lem:bias}
Consider the setting of Sec.~\ref{sec: diff of exp}. Take $g$ as in Theorem.~\ref{rq:thm:main}. We then have
\begin{equation}
\begin{aligned}
 \tilde{b}_P&=\rqnorm*{\rqE[P(\tilde{H}_\ell)-P(\tilde{\bar{H}})]}\\
 &\leq\frac{\tilde{\nu}}{2\sqrt{2\pi}}\rqnorm{\widehat{g^\prime}}_1.
\end{aligned}
\label{rq:eq:bP}
\end{equation}
\end{lemma}

\begin{proof}
Expand $\tilde{H}_\ell=\tilde{\bar{H}}+\tilde{E}_\ell$. Recall from Eq.~\eqref{rq:eq:delta-integral} in the proof of
Lemma~\ref{rq:lem:variance} that
\begin{equation}
\begin{aligned}
 \tilde{\Delta}_\ell
 &=P(\tilde{\bar{H}}+\tilde{E}_\ell)-P(\tilde{\bar{H}})\\
 &=\frac{1}{\sqrt{2\pi}}\int_{\mathbb{R}}\frac{\widehat{g}(t)}{it}
    \left(e^{it(\tilde{\bar{H}}+\tilde{E}_\ell)}-e^{it\tilde{\bar{H}}}\right)dt\\
\end{aligned}
\label{rq:eq:bias-start}
\end{equation}
For fixed $t\in\mathbb{R}$, define $f_t(X)=e^{itX}$.
We can expand $f_t(\tilde{\bar{H}})$ in the direction $\tilde{E}_\ell$ as~\cite{higham_higher_2014}:
\begin{equation}
\begin{aligned}
 e^{it(\tilde{\bar{H}}+\tilde{E}_\ell)}
 &=e^{it\tilde{\bar{H}}}+\rqFrechet{f_t}{\tilde{\bar{H}}}{\tilde{E}_\ell}\\
 &\quad+R_{\tilde{\bar{H}}}(t;\tilde{E}_\ell).
\end{aligned}
\label{rq:eq:frechet-expansion}
\end{equation}Here
\begin{equation}
 \rqFrechet{f_t}{\tilde{\bar{H}}}{\tilde{E}_\ell}
 =\left.\frac{d}{d\eta}e^{it(\tilde{\bar{H}}+\eta \tilde{E}_\ell)}
   \right|_{\eta=0}
\label{rq:eq:frechet-definition}
\end{equation}is the Fr\'{e}chet derivative in the direction of $\tilde{E}_\ell$ and
$R_{\tilde{\bar{H}}}(t;\tilde{E}_\ell)$ is the remainder satisfying $\rqnorm{R_{\tilde{\bar{H}}}(t;\tilde{E}_\ell)} = o(\rqnorm{\tilde{E}_\ell})$.~\cite{higham_higher_2014} We have
\begin{equation}
\begin{aligned}
 \rqFrechet{f_t}{\tilde{\bar{H}}}{\tilde{E}_\ell}
 &=\lim_{\eta\to0}
   \frac{e^{it(\tilde{\bar{H}}+\eta \tilde{E}_\ell)}-e^{it\tilde{\bar{H}}}}{\eta}\\
 &=\lim_{\eta\to0}it\int_0^1
   e^{it(1-u)(\tilde{\bar{H}}+\eta \tilde{E}_\ell)}
   \tilde{E}_\ell e^{itu\tilde{\bar{H}}}\,du\\
 &=it\int_0^1 e^{it(1-u)\tilde{\bar{H}}}
   \tilde{E}_\ell e^{itu\tilde{\bar{H}}}\,du.
\end{aligned}
\label{rq:eq:frechet-duhamel}
\end{equation}
Thus
\begin{equation}
\begin{aligned}
 \rqE[\rqFrechet{f_t}{\tilde{\bar{H}}}{\tilde{E}_\ell}]
 &=it\int_0^1e^{it(1-u)\tilde{\bar{H}}}
   \rqE[\tilde{E}_\ell]e^{itu\tilde{\bar{H}}}\,du\\
 &=0,
\end{aligned}
\label{rq:eq:linear-cancellation}
\end{equation}
where we used $\rqE[\tilde{E}_\ell]=0$. Consequently,
\begin{equation}
\begin{aligned}
 \rqE[\tilde{\Delta}_\ell]
 &=\frac{1}{\sqrt{2\pi}}\int_{\mathbb{R}}\frac{\widehat{g}(t)}{it}
   \rqE\!\left[e^{it(\tilde{\bar{H}}+\tilde{E}_\ell)}-e^{it\tilde{\bar{H}}}\right]dt\\
 &=\frac{1}{\sqrt{2\pi}}\int_{\mathbb{R}}\frac{\widehat{g}(t)}{it}
   \rqE\!\left[\rqFrechet{f_t}{\tilde{\bar{H}}}{\tilde{E}_\ell}
          +R_{\tilde{\bar{H}}}(t;\tilde{E}_\ell)\right]dt\\
 &=\frac{1}{\sqrt{2\pi}}\int_{\mathbb{R}}\frac{\widehat{g}(t)}{it}
   \rqE[R_{\tilde{\bar{H}}}(t;\tilde{E}_\ell)]\,dt.
\end{aligned}
\label{rq:eq:mean-remainder}
\end{equation}Next we compute the remainder. First,
\begin{equation}
\begin{aligned}
 R_{\tilde{\bar{H}}}(t;\tilde{E}_\ell)
 &=e^{it(\tilde{\bar{H}}+\tilde{E}_\ell)}-e^{it\tilde{\bar{H}}}
   -\rqFrechet{f_t}{\tilde{\bar{H}}}{\tilde{E}_\ell}\\
 &=-\left[e^{it\tilde{\bar{H}}}-e^{it(\tilde{\bar{H}}+\tilde{E}_\ell)}\right]
   -\rqFrechet{f_t}{\tilde{\bar{H}}}{\tilde{E}_\ell}\\
 &=-it\int_0^1 e^{it(1-u)\tilde{\bar{H}}}(-\tilde{E}_\ell)
     e^{itu(\tilde{\bar{H}}+\tilde{E}_\ell)}\,du\\
 &\qquad-\rqFrechet{f_t}{\tilde{\bar{H}}}{\tilde{E}_\ell}.
\end{aligned}
\label{rq:eq:remainder-start}
\end{equation}
Adding and subtracting $e^{itu\tilde{\bar{H}}}$ inside the integrand gives
\begin{equation}
\begin{aligned}
 R_{\tilde{\bar{H}}}(t;\tilde{E}_\ell)
 &=it\int_0^1 e^{it(1-u)\tilde{\bar{H}}}\tilde{E}_\ell
     \Bigl[e^{itu(\tilde{\bar{H}}+\tilde{E}_\ell)}\\
 &\hspace{71pt}+e^{itu\tilde{\bar{H}}}-e^{itu\tilde{\bar{H}}}\Bigr]du\\
 &\qquad-\rqFrechet{f_t}{\tilde{\bar{H}}}{\tilde{E}_\ell}\\
 &=it\int_0^1 e^{it(1-u)\tilde{\bar{H}}}\tilde{E}_\ell
     \Bigl[e^{itu(\tilde{\bar{H}}+\tilde{E}_\ell)}-e^{itu\tilde{\bar{H}}}\Bigr]du.
\end{aligned}
\label{rq:eq:remainder-single-integral}
\end{equation}

To evaluate the difference in the last line, use
Lemma~\ref{rq:lem:duhamel} once more:
\begin{equation}
\begin{aligned}
 &e^{itu(\tilde{\bar{H}}+\tilde{E}_\ell)}-e^{itu\tilde{\bar{H}}}\\
 &\quad=itu\int_0^1 e^{itu(1-s)(\tilde{\bar{H}}+\tilde{E}_\ell)}
   \tilde{E}_\ell e^{itus\tilde{\bar{H}}}\,ds.
\end{aligned}
\label{rq:eq:inner-duhamel}
\end{equation}Substituting $w=us$ yields
\begin{equation}
\begin{aligned}
 &e^{itu(\tilde{\bar{H}}+\tilde{E}_\ell)}-e^{itu\tilde{\bar{H}}}\\
 &\quad=it\int_0^u e^{it(u-w)(\tilde{\bar{H}}+\tilde{E}_\ell)}
   \tilde{E}_\ell e^{itw\tilde{\bar{H}}}\,dw.
\end{aligned}
\label{rq:eq:inner-duhamel-rescaled}
\end{equation}
Therefore
\begin{multline}
 \rqE[R_{\tilde{\bar{H}}}(t;\tilde{E}_\ell)]
 =(it)^2\int_0^1\int_0^u e^{it(1-u)\tilde{\bar{H}}}\\
 \times\rqE\!\left[
   \tilde{E}_\ell e^{it(u-w)(\tilde{\bar{H}}+\tilde{E}_\ell)}\tilde{E}_\ell
 \right]e^{itw\tilde{\bar{H}}}\,dw\,du.
\label{rq:eq:remainder-double-integral}
\end{multline}
By Lemma~\ref{rq:lem:block},
\begin{equation}
 \rqnorm*{\rqE\!\left[
   \tilde{E}_\ell e^{it(u-w)(\tilde{\bar{H}}+\tilde{E}_\ell)}\tilde{E}_\ell
 \right]}\leq\tilde{\nu}.
\label{rq:eq:remainder-integrand-bound}
\end{equation}
Thus
\begin{equation}
\begin{aligned}
 \rqnorm*{\rqE[R_{\tilde{\bar{H}}}(t;\tilde{E}_\ell)]}
 &\leq t^2\tilde{\nu}\int_0^1\int_0^u dw\,du\\
 &=\frac{t^2\tilde{\nu}}{2}.
\end{aligned}
\label{rq:eq:remainder-bound}
\end{equation}
Recalling Eq.~\eqref{rq:eq:mean-remainder}, we obtain
\begin{equation}
\begin{aligned}
 \rqnorm*{\rqE[\tilde{\Delta}_\ell]}
 &\leq\frac{1}{\sqrt{2\pi}}\int_{\mathbb{R}}
   \rqabs*{\frac{\widehat{g}(t)}{t}}
   \rqnorm*{\rqE[R_{\tilde{\bar{H}}}(t;\tilde{E}_\ell)]}\,dt\\
 &\leq\frac{\tilde{\nu}}{2\sqrt{2\pi}}\int_{\mathbb{R}}
       \rqabs{t\widehat{g}(t)}\,dt\\
 &=\frac{\tilde{\nu}}{2\sqrt{2\pi}}\rqnorm{t\widehat{g}}_1\\
 &=\frac{\tilde{\nu}}{2\sqrt{2\pi}}\rqnorm{\widehat{g^\prime}}_1
\end{aligned}
\label{rq:eq:bias-Bg}
\end{equation}where in the last line we have used the fact that $it\widehat{g}(t) = \widehat{g^\prime}(t)$, which can be shown by integration by parts.
\end{proof}

\subsection{Bounds on $b_P$ and $q_P$ for non-hermitian operators}
    Here we offer a simple extension of the bounds on $b_P$ and $q_P$ for non-hermitian operators.
\begin{lemma}[Extension to non-hermitian operators]\label{lem: extended lemma}
Let $\left\{p_\ell, A_\ell\right\}$ be a finite ensemble of linear operators $A_\ell\in\mathbb{C}^{m\times n}$. Consider the set-up of Theorem~\ref{rq:thm:main} and the definitions in Eq.~\eqref{rq:eq:errors}. Then:
\begin{equation}
    q_P\leq \frac{\nu}{2\pi} \rqnorm{\widehat{g}}_1^2, \qquad b_P \leq \frac{\nu}{2\sqrt{2\pi}}\rqnorm{\widehat{g^\prime}}_1
\end{equation}

\begin{proof}
    Consider the hermitian dilation of a linear operator $X\in \mathbb{C}^{m\times n}$:
    \begin{equation}
        \mathcal{H}(X) = \begin{pmatrix}
            0 & X\\
            X^\dagger & 0\\
        \end{pmatrix}.
    \end{equation}Then by (the full arXiv version of) Ref.~\cite{gilyen_quantum_2019}, Eqs. 34, 35 we have
    \begin{equation}\label{eq: P odd dilation}
        P(\mathcal{H}(X)) = \begin{pmatrix}
            0 & P^{\sv}(X)\\
            P^{\sv}(X^\dagger) & 0
        \end{pmatrix}
    \end{equation}for odd $P$ and
    \begin{equation}\label{eq: P even dilation}
        P(\mathcal{H}(X)) = \begin{pmatrix}
            P^{\sv}(X^\dagger) & 0\\
            0 & P^{\sv}(X)
        \end{pmatrix}
    \end{equation}for even $P$. Define $\tilde{H}_\ell = \mathcal{H}(A_\ell)$, $\tilde{\bar{H}} = \mathcal{H}(\bar{A})$, and $\tilde{E}_\ell = \mathcal{H}(E_\ell)$. Then
    \begin{equation}
    \tilde{V} = \mathbb{E}\left[\tilde{E}_\ell^2\right] = \begin{pmatrix}
        V_L & 0\\
        0 & V_R
    \end{pmatrix}, \qquad \rqnorm{\tilde{V}} = \tilde{\nu}.
    \end{equation}Let $\tilde{\Delta}_\ell = P(\tilde{H}_\ell) - P(\tilde{\bar{H}})$. By Lemmas~\ref{rq:lem:variance} and~\ref{rq:lem:bias} we have:
    \begin{equation}
    \begin{split}
        &\tilde{q}_P = \left\|\mathbb{E}\left[\tilde{\Delta}_\ell^\dagger\tilde{\Delta}_\ell\right]\right\| \leq \frac{\tilde{\nu}}{2\pi} \rqnorm{\widehat{g}}_1^2\\
        &\tilde{b}_P = \left\|\mathbb{E}\left[\tilde{\Delta}_\ell\right]\right\| \leq \frac{\tilde{\nu}}{2\sqrt{2\pi}}\rqnorm{\widehat{g^\prime}}_1.     
    \end{split}
    \end{equation}All that remains is to prove that $q_P \leq \tilde{q}_P$ and $b_P\leq \tilde{b}_P$. By inspection of Eqs.~\eqref{eq: P odd dilation} and~\eqref{eq: P even dilation} we can see that $\rqnorm{\mathbb{E}[\Delta_\ell]}\leq \rqnorm{\mathbb{E}[\tilde
    {\Delta}_\ell]}$. Thus, $b_P \leq \tilde{b}_P$. Define:
    \begin{equation}
        \begin{split}
            J_0 = \begin{pmatrix}
                I\\
                0
            \end{pmatrix}, \qquad J_1 = \begin{pmatrix}
                0\\
                I
            \end{pmatrix}.
        \end{split}
    \end{equation}Further define $J_\mathrm{out} = J_1$ for even $P$ and $J_{\mathrm{out}} = J_0$ for odd $P$. Then we can write $\Delta_\ell = J_{\mathrm{out}}^\dagger\tilde{\Delta}_\ell J_1$. Thus:
    \begin{equation}
        \mathbb{E}\left[\Delta_\ell^\dagger \Delta_\ell\right] = J_1^\dagger\mathbb{E}\left[\tilde{\Delta}_\ell^\dagger J_{\mathrm{out}}J_{\mathrm{out}}^\dagger\tilde{\Delta}_\ell\right]J_1.
    \end{equation}For any vector $z \in \mathbb{C}^{m+n}$, we have:
    \begin{equation}
        \begin{split}
        &z^\dagger \tilde{\Delta}_\ell^\dagger(I_{m+n} - J_{\mathrm{out}}J_{\mathrm{out}}^\dagger)\tilde{\Delta}_\ell z\\
        &= (\tilde{\Delta}_\ell z)^\dagger(I_{m+n} - J_{\mathrm{out}}J_{\mathrm{out}}^\dagger)(\tilde{\Delta}_\ell z)\\
        &\geq 0
        \end{split}
    \end{equation}which follows from the fact that
    \begin{equation}
        I_{m+n} - J_{\mathrm{out}}J_{\mathrm{out}}^\dagger = \begin{cases}
            \mathrm{diag}(I, 0) & \text{if $P$ is even}\\
        \mathrm{diag}(0, I) & \text{if $P$ is odd}
        \end{cases}
    \end{equation}is positive semi-definite. It follows that
    \begin{equation}
        \left\|\mathbb{E}\left[\tilde{\Delta}_\ell^\dagger J_{\mathrm{out}}J_{\mathrm{out}}^\dagger\tilde{\Delta}_\ell\right]\right\|\leq \left\|\mathbb{E}\left[\tilde{\Delta}_\ell^\dagger \tilde{\Delta}_\ell\right]\right\|.
    \end{equation}Thus
    \begin{equation}
        \begin{split}
            &\left\|J_1^\dagger\mathbb{E}\left[\tilde{\Delta}_\ell^\dagger J_{\mathrm{out}}J_{\mathrm{out}}^\dagger\tilde{\Delta}_\ell\right]J_1\right\|\\
            &\leq \rqnorm{J_1^\dagger} \left\|\mathbb{E}\left[\tilde{\Delta}_\ell^\dagger J_{\mathrm{out}}J_{\mathrm{out}}^\dagger\tilde{\Delta}_\ell\right]\right\|\rqnorm{J_1}\\
            &\leq \left\|\mathbb{E}\left[\tilde{\Delta}_\ell^\dagger \tilde{\Delta}_\ell\right]\right\|.
        \end{split}
    \end{equation}Thus, $q_P \leq \tilde{q}_P$.
\end{proof}
    
\end{lemma}

\section{Prefactors for common function families}\label{sec: prefactor calculations}

We use the Fourier transform convention:
\begin{equation}
\hat{g}(t)
=\frac{1}{\sqrt{2\pi}}
\int_{\mathbb{R}} g(x)e^{-itx}\,dx.
\end{equation}We will also frequently make use of the identity from Eq. 5.9.3 of Ref.~\cite{dlmf_nodate}:
\begin{equation}
    \int_{\mathbb{R}}|t|^{2z-1}e^{-ct^2}dt = c^{-z}\Gamma (z)
\end{equation}where $\Gamma(z)$ is the gamma function. We now explicitly compute $C_f$ for cosine, erf, and Gaussian functions.

\subsection{Cosine prefactor}

Let $f(x)=\cos(kx)$, $g(x)=-k\sin(kx)$. We write:
\begin{equation}
\begin{aligned}
\cos(kx)&=\frac{e^{ikx}+e^{-ikx}}{2}
\\
\sin(kx)&=\frac{e^{ikx}-e^{-ikx}}{2i}
\end{aligned}
\end{equation}
which allows us to write:
\begin{equation}
\begin{aligned}
\hat{g}(t)
&=\frac{-k}{2i\sqrt{2\pi}}
\int_{\mathbb{R}}
\bigl(e^{ix(k-t)}-e^{-ix(k+t)}\bigr)\,dx
\\
&=-\frac{k\sqrt{2\pi}}{2i}
\bigl(\delta(k-t)-\delta(k+t)\bigr)
\end{aligned}
\end{equation}
\begin{equation}
\begin{aligned}
\widehat{g'}(t)
&=\frac{-k^2}{2\sqrt{2\pi}}
\int_{\mathbb{R}}
\bigl(e^{ix(k-t)}+e^{-ix(k+t)}\bigr)\,dx
\\
&=-\frac{k^2\sqrt{2\pi}}{2}
\bigl(\delta(k-t)+\delta(k+t)\bigr).
\end{aligned}
\end{equation}

Thus:
\begin{equation}
\begin{aligned}
\|\hat{g}\|_1
&=\frac{k\sqrt{2\pi}}{2}
\int_{\mathbb{R}}
\bigl(\delta(k-t)+\delta(k+t)\bigr)\,dt
\\
&=k\sqrt{2\pi}
\end{aligned}
\end{equation}
\begin{equation}
\begin{aligned}
\|\widehat{g'}\|_1
&=\frac{k^2\sqrt{2\pi}}{2}
\int_{\mathbb{R}}
\bigl(\delta(k-t)+\delta(k+t)\bigr)\,dt
\\
&=k^2\sqrt{2\pi}.
\end{aligned}
\end{equation}

Thus:
\begin{equation}
C_f=2k^2.
\end{equation}

\subsection{Erf prefactor}

Let $f(x)=\operatorname{erf}(kx)$,
$g(x)=\frac{2k}{\sqrt{\pi}}e^{-k^2x^2}$. Then:
\begin{equation}
\begin{aligned}
\hat{g}(t)
&=\frac{1}{\sqrt{2\pi}}
\int_{\mathbb{R}}
\frac{2k}{\sqrt{\pi}}e^{-k^2x^2}e^{-itx}\,dx
\\
&=\frac{\sqrt{2}k}{\pi}
\int_{\mathbb{R}}e^{-k^2x^2}e^{-itx}\,dx
\\
&=\sqrt{\frac{2}{\pi}}e^{-t^2/4k^2}
\end{aligned}
\end{equation}
\begin{equation}
\begin{aligned}
\widehat{g'}(t)
&=it\hat{g}(t)
\\
&=i\sqrt{\frac{2}{\pi}}\,t\,e^{-t^2/4k^2}
\end{aligned}
\end{equation}

This gives us:
\begin{equation}
\begin{aligned}
\|\hat{g}\|_1
&=\sqrt{\frac{2}{\pi}}
\int_{\mathbb{R}}e^{-t^2/4k^2}\,dt
\\
&=\sqrt{\frac{2}{\pi}}(2k)
\Gamma\left(\frac{1}{2}\right)
\\
&=2\sqrt{2}\,k
\end{aligned}
\end{equation}
\begin{equation}
\begin{aligned}
\|\widehat{g'}\|_1
&=\sqrt{\frac{2}{\pi}}
\int_{\mathbb{R}}|t|e^{-t^2/4k^2}\,dt
\\
&=\sqrt{\frac{2}{\pi}}(4k^2)\Gamma(1)
\\
&=4\sqrt{\frac{2}{\pi}}k^2.
\end{aligned}
\end{equation}

Thus:
\begin{equation}
C_f=\frac{4}{\pi}\left(1+\mathrm{erf}(k)\right)k^2.
\end{equation}

\subsection{Gaussian prefactor}

Let $f(x)=e^{-\beta x^2}$, $g(x)=f'(x)$. Use the fact that $it\hat{f}(t)=\widehat{f'}(t)$,
$-t^2\hat{f}(t)=\widehat{f''}(t)$.

Then:
\begin{equation}
\begin{aligned}
\hat{g}(t)
&=\frac{it}{\sqrt{2\pi}}
\int_{\mathbb{R}}e^{-\beta x^2}e^{-ixt}\,dx
\\
&=\frac{it}{\sqrt{2\beta}}e^{-t^2/4\beta}
\end{aligned}
\end{equation}
\begin{equation}
\begin{aligned}
\widehat{g'}(t)
&=\frac{-t^2}{\sqrt{2\pi}}
\int_{\mathbb{R}}e^{-\beta x^2}e^{-ixt}\,dx
\\
&=\frac{-t^2}{\sqrt{2\beta}}e^{-t^2/4\beta}
\end{aligned}
\end{equation}

Thus:
\begin{equation}
\begin{aligned}
\|\hat{g}\|_1
&=\frac{1}{\sqrt{2\beta}}
\int_{\mathbb{R}}|t|e^{-t^2/4\beta}\,dt
\\
&=\frac{1}{\sqrt{2\beta}}(4\beta)\Gamma(1)
\\
&=\frac{4\sqrt{\beta}}{\sqrt{2}}
\end{aligned}
\end{equation}
\begin{equation}
\begin{aligned}
\|\widehat{g'}\|_1
&=\frac{1}{\sqrt{2\beta}}
\int_{\mathbb{R}}t^2e^{-t^2/4\beta}\,dt
\\
&=\frac{1}{\sqrt{2\beta}}(4\beta)^{3/2}
\Gamma\left(\frac{3}{2}\right)
\\
&=2\sqrt{2\pi}\,\beta
\end{aligned}
\end{equation}

Thus:
\begin{equation}
C_f=\left(\frac{4}{\pi}+2\right)\beta.
\end{equation}

\section{Example of $P^\prime$ extension}\label{sec: g construction}

On the interval $(1,w+1)$ for $w > 0$, let $x$ and $s$ be related by:
\begin{equation}
x=1+ws,\qquad s\in(0,1)
\end{equation}
and on the interval $(-w-1,-1)$:
\begin{equation}
x=-1-ws,\qquad s\in(0,1).
\end{equation}

Define:
\begin{equation}
\begin{aligned}
h_0(s)&=1-3s^2+2s^3,\\
h_1(s)&=s-2s^2+s^3.
\end{aligned}
\end{equation}We can then construct a valid extension function $g(x)$ as:
\begin{equation}
g(x)=
\begin{cases}
P'(x), & x\in[-1,1]\\
P'(1)h_0(s)+wP''(1)h_1(s), & x\in(1,w+1)\\
P'(-1)h_0(s)-wP''(-1)h_1(s), & x\in(-w-1,-1)\\
0, & |x|\ge w+1
\end{cases}
\end{equation}
\begin{equation}
\begin{aligned}
\frac{dh_0}{ds}&=-6s+6s^2\\
\frac{dh_1}{ds}&=1-4s+3s^2
\end{aligned}
\end{equation}On $x\in(1,w+1)$:
\begin{equation}
\begin{aligned}
\frac{dg}{dx}
&=P'(1)\frac{dh_0}{ds}\frac{ds}{dx}
  +wP''(1)\frac{dh_1}{ds}\frac{ds}{dx}\\
&=P'(1)(-6s+6s^2)\frac{1}{w}\\
&\quad{}+P''(1)(1-4s+3s^2)
\end{aligned}
\end{equation}
\begin{equation}
\begin{aligned}
\frac{d^2g}{dx^2}
&=\frac{1}{w^2}P'(1)(-6+12s)\\
&\quad{}+\frac{1}{w}P''(1)(-4+6s).
\end{aligned}
\end{equation}On $x\in(-w-1,-1)$:
\begin{equation}
\begin{aligned}
\frac{dg}{dx}
&=P'(-1)\frac{dh_0}{ds}\frac{ds}{dx}\\
&\quad{}-wP''(-1)\frac{dh_1}{ds}\frac{ds}{dx}\\
&=-P'(-1)(-6s+6s^2)\frac{1}{w}\\
&\quad{}+P''(-1)(1-4s+3s^2)
\end{aligned}
\end{equation}
\begin{equation}
\begin{aligned}
\frac{d^2g}{dx^2}
&=\frac{1}{w^2}P'(-1)(-6+12s)\\
&\quad{}-\frac{1}{w}P''(-1)(-4+6s).
\end{aligned}
\end{equation}This function construction can be viewed as an interpolation between the pairs of points $x = 1,1+w$ and $x = -w-1,-1$ using cubic Hermite functions, similar to what has been done in the context of data fitting~\cite{cubic_interpolation}.

\section{Additional Error Bound Plots}\label{sec: additional error bound plots}

\begin{figure*}[htb!]
    \centering
    \includegraphics[width=0.95\linewidth]{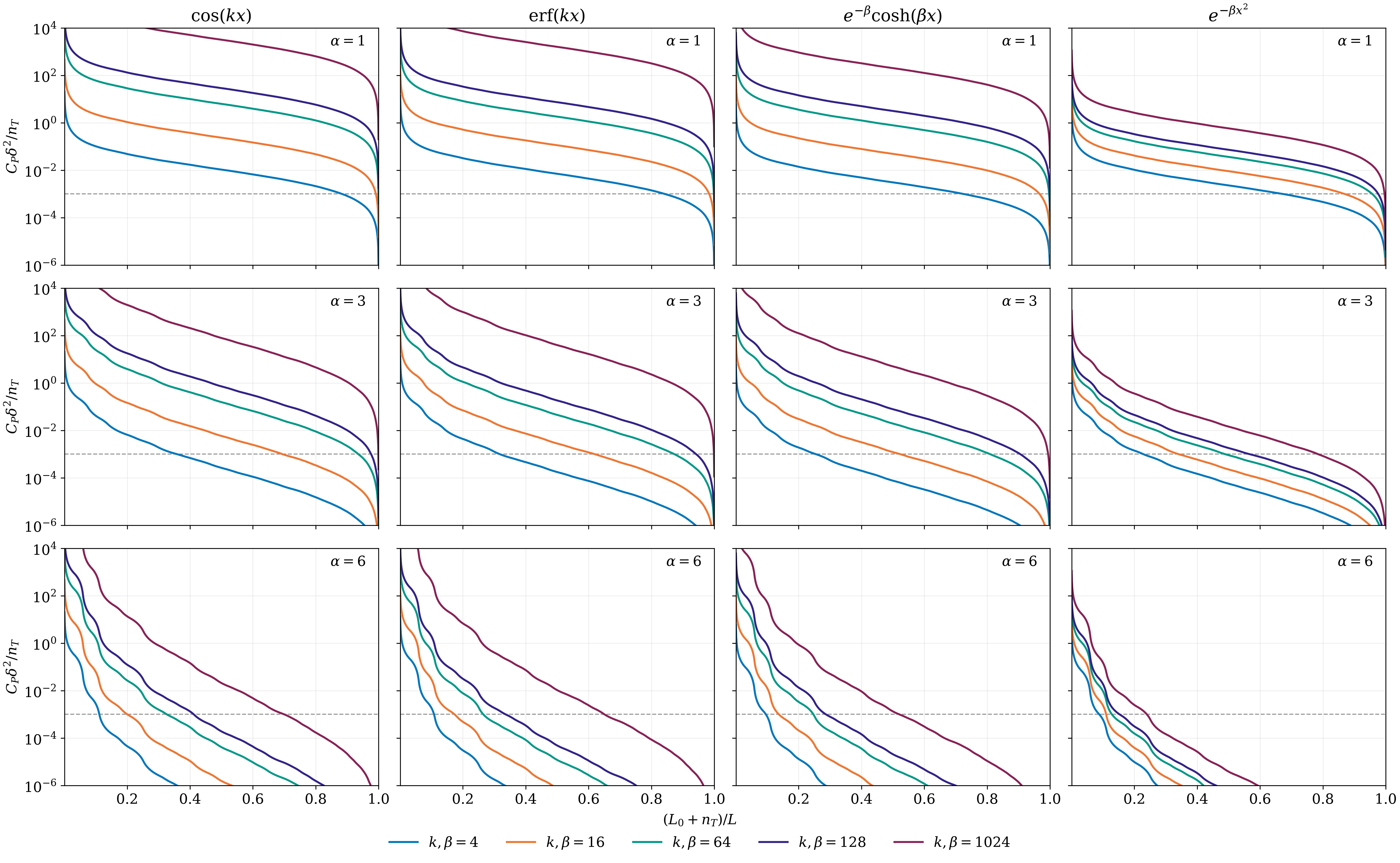}
    \caption{The diamond norm error bound in Thm.~\ref{rq:thm:main} as a function of ($L_0+n_T$)/L for the long-range Heisenberg model in Eq.~\eqref{eq: long-range heisenberg} for cosine, error, hyperbolic cosine, and Gaussian filtering functions. Here, $\delta = \lambda_\epsilon/\lambda$. The value of $\alpha$ governs the power-law of the long-range interactions. The horizontal grey dashed line denotes a target error bound of $10^{-3}$.}
    \label{fig:heisenberg_plots_full}
\end{figure*}

\begin{figure*}[htb!]
    \centering
    \includegraphics[width=0.95\linewidth]{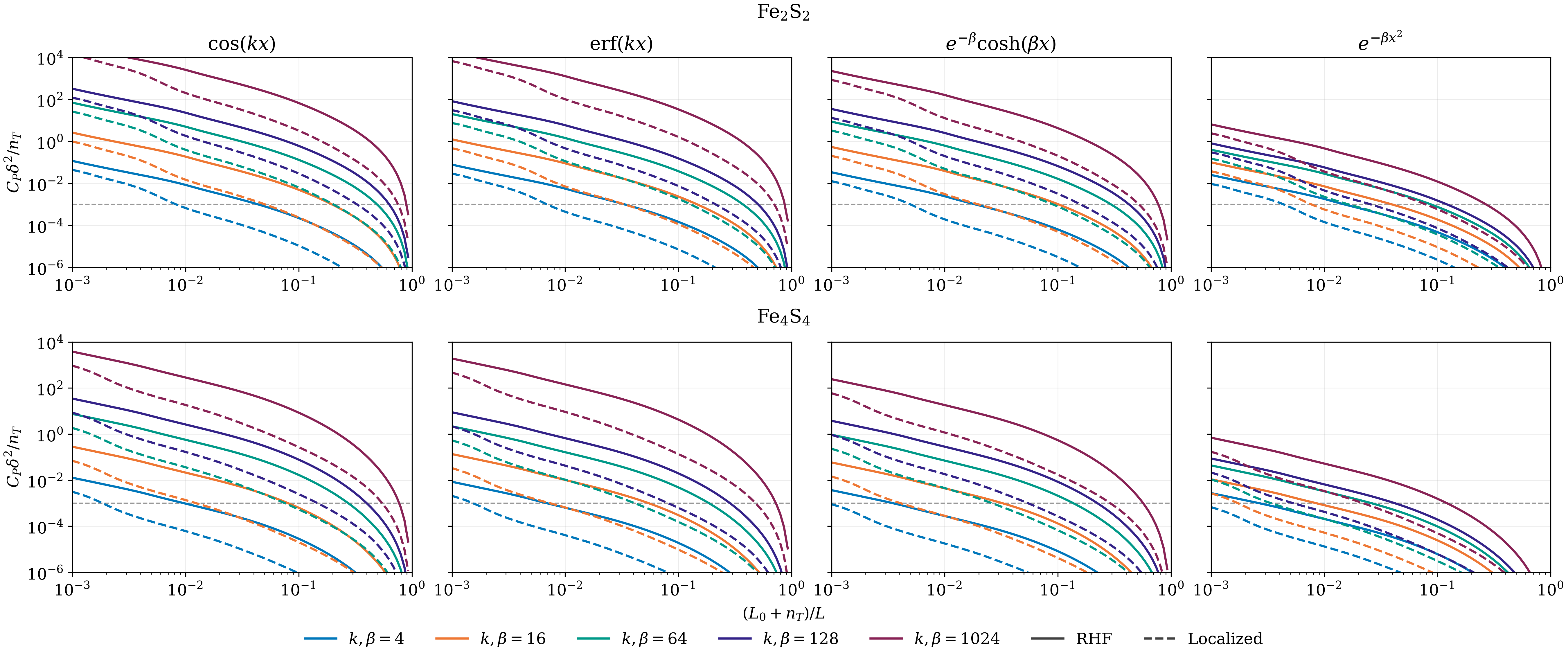}
    \caption{The diamond norm error bounds from Thm.~\ref{rq:thm:main} as a function of the fractional Pauli term budget for the iron-sulfur complexes $\mathrm{Fe_2S_2}$ and $\mathrm{Fe_4S_4}$ for cosine, error, hyperbolic cosine, and Gaussian filtering functions. Dashed lines denote the localized DFT orbitals described in Ref.~\cite{li_chan_fcidump} and solid lines denote the transformed MO orbitals from Ref.~\cite{robledo-moreno_chemistry_2025} in the same active space.}
    \label{fig:iron_sulfur_plots_full}
\end{figure*}

In this section we provide additional plots that were omitted from the main text for the sake of conciseness. In particular, we include plots for the $\alpha = 1$ long-range Heisenberg model as well as plots for the error and hyperbolic cosine functions, which are relevant to filtering and Gibbs state preparation, respectively. In Fig.~\ref{fig:heisenberg_plots_full} we can see that the results for the $\alpha =1$ long-range Heisenberg model are qualitatively similar to the $\alpha = 3$ results, but are quantitatively worse for RBE-QSVT than $\alpha = 3,6$. In Fig.~\ref{fig:iron_sulfur_plots_full} we add the error and hyperbolic cosine plots to the target functions studied for the iron-sulfur complexes, which we can see are qualitatively similar to the error bounds for the cosine and Gaussian filtering functions presented in the main text.

\end{document}